\documentclass[lettersize,journal]{IEEEtran}
\usepackage{amsmath,amsfonts}
\usepackage{algorithmic}
\usepackage{algorithm}
\usepackage{array}
\usepackage[caption=false,font=footnotesize,labelfont=rm,textfont=rm]{subfig}
\usepackage{textcomp}
\usepackage{stfloats}
\usepackage{url}
\usepackage{verbatim}
\usepackage{graphicx}
\usepackage{cite}
\usepackage{threeparttable}
\usepackage{tikz}
\usepackage{colortbl,xcolor,array}
\usepackage{multicol}
\usepackage{multirow}
\usepackage{float}
\usepackage{booktabs}
\usepackage[switch]{lineno}

\begin{document}
\title{Spiking Tucker Fusion Transformer for Audio-Visual Zero-Shot Learning}

\author{Wenrui Li,
        Penghong Wang,
        Ruiqin Xiong,~\IEEEmembership{~Senior Member,~IEEE,}
        Xiaopeng Fan,~\IEEEmembership{~Senior Member,~IEEE}
        
\thanks{This work was supported in part by the National Key R\&D Program of China (2021YFF0900500), and the National Natural Science Foundation of China (NSFC) under grants 62441202, U22B2035. (Corresponding author: Xiaopeng Fan.)}
\thanks{Wenrui Li, Penghong Wang and Xiaopeng Fan are with the Department of Computer Science and Technology, Harbin Institute of Technology, Harbin 150001, China. (e-mail: liwr@stu.hit.edu.cn; phwang@hit.edu.cn; fxp@hit.edu.cn).}
\thanks{Ruiqin Xiong is with the School of Electronic Engineering and Computer Science, Institute of Digital Media, Peking University, Beijing 100871, China (e-mail: rqxiong@pku.edu.cn)}
}



\maketitle

\begin{abstract}
The spiking neural networks (SNNs) that efficiently encode temporal sequences have shown great potential in extracting audio-visual joint feature representations. However, coupling SNNs (binary spike sequences) with transformers (float-point sequences) to jointly explore the temporal-semantic information still facing challenges. In this paper, we introduce a novel Spiking Tucker Fusion Transformer (STFT) for audio-visual zero-shot learning (ZSL). The STFT leverage the temporal and semantic information from different time steps to generate robust representations. The time-step factor (TSF) is introduced to dynamically synthesis the subsequent inference information. To guide the formation of input membrane potentials and reduce the spike noise, we propose a global-local pooling (GLP) which combines the max and average pooling operations. Furthermore, the thresholds of the spiking neurons are dynamically adjusted based on semantic and temporal cues. Integrating the temporal and semantic information extracted by SNNs and Transformers are difficult due to the increased number of parameters in a straightforward bilinear model. To address this, we introduce a temporal-semantic Tucker fusion module, which achieves multi-scale fusion of SNN and Transformer outputs while maintaining full second-order interactions. Our experimental results demonstrate the effectiveness of the proposed approach in achieving state-of-the-art performance in three benchmark datasets. The harmonic mean (HM) improvement of VGGSound, UCF101 and ActivityNet are around 15.4\%, 3.9\%, and 14.9\%, respectively.
\end{abstract}


\section{Introduction}
\IEEEPARstart{T}{he} task of audio-visual zero-shot learning (ZSL) \cite{avgzslnet,avca,tcaf} aims to classify objects or scenes by utilizing both audio and visual modalities, even when labeled data is not available. Conventional supervised audio-visual approaches are training with lots of labeled training instances for each class. In order to address the constraints of traditional supervised audio-visual methods, the generalized zero-shot learning (GZSL) setting has been proposed \cite{hong2023hyperbolic,li2023modality}. GZSL methods permit models to identify and classify instances from both seen and unseen classes, thereby facilitating more practical and scalable solutions for audio-visual classification and recognition tasks.
\begin{figure}
    \centering
	\includegraphics[scale=0.42]{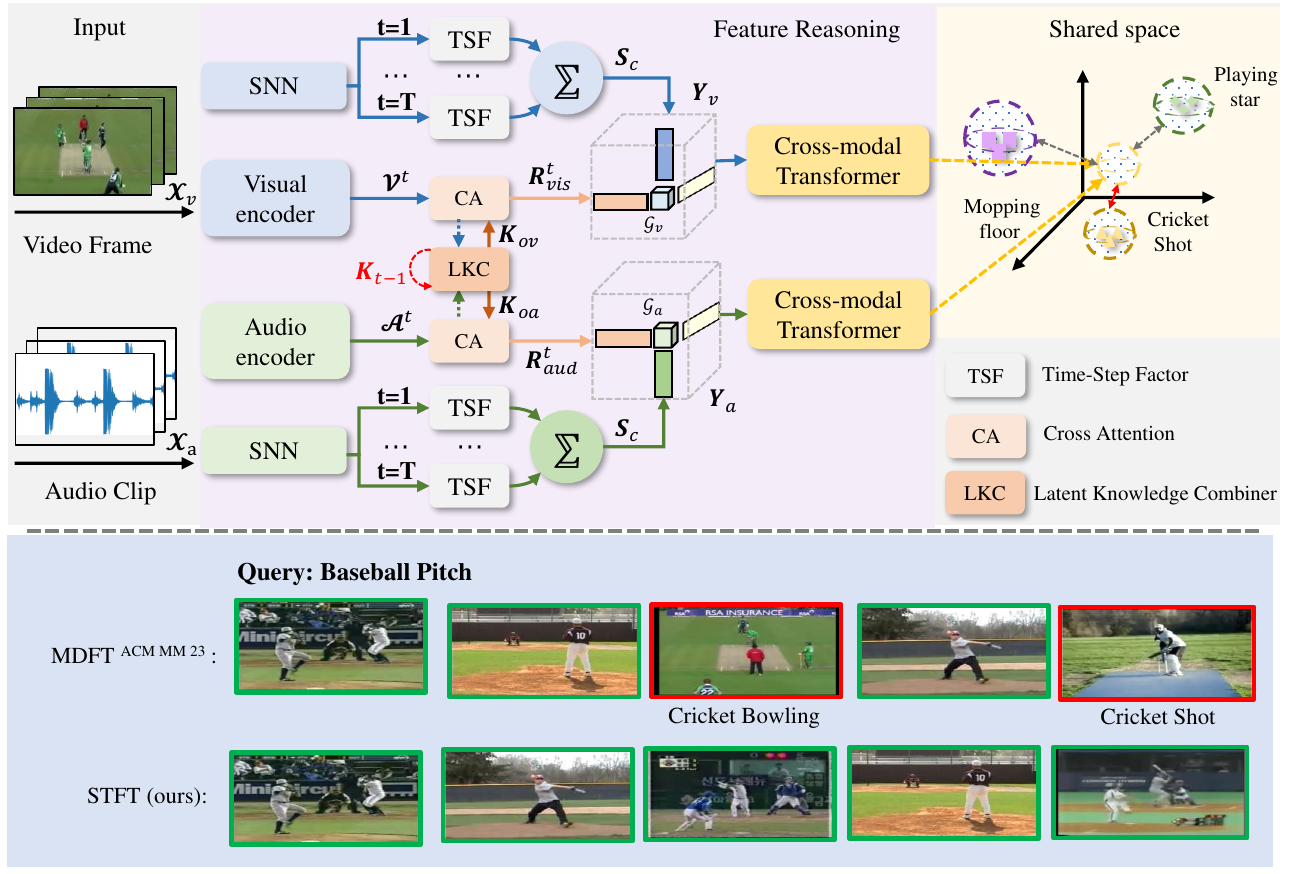}
	\caption{The illustration of our proposed STFT for audio-visual GZSL. The SNN utilize the time-step factor to dynamic synthesis the output of the temporal information. The audio and visual encoder utilize the latent knowledge combiner to explore the semantic information with latent cues. After temporal-semantic tucker fusion, the fused features are further reasoned through the cross-modal transformer. The information from seen training classes could transfer to unseen test classes by textual embeddings.}
	\label{fig1}
\end{figure}

To obtain more robust audio-visual feature representations, most existing methods model and align the temporal and semantic features of the input separately. CJME \cite{cjme} projects the audio-visual and textual modalities into a shared space and calculates their similarity, clustering features of the same category in the shared space using triplet loss. Mercea et al. \cite{avca} introduce a lightweight processing framework that achieves excellent results by utilizing cross-attention to interact with audio-visual modality information. TCaF \cite{tcaf} preprocesses temporal information and verifies its importance in the interaction of audio-visual modalities. Spike Neural Networks (SNNs) provide significant advantages for audio-visual representation. Firstly, they efficiently encode temporal information by mimicking the spike-timing of biological neurons, allowing precise modeling of dynamic events over time. This spike-timing-dependent plasticity (STDP) enables SNNs to capture fine-grained temporal patterns crucial for understanding complex audio-visual data. Secondly, SNNs offer high stability and robustness to noise, making them resilient to variations and disturbances in real-world data. This robustness is particularly beneficial in scenarios where audio and visual inputs are degraded or incomplete. Thirdly, integrating SNNs with transformers enhances the extraction of both temporal and semantic features. SNNs manage precise timing aspects, while transformers excel at capturing contextual relationships, resulting in a comprehensive multimodal feature representation. This combination has demonstrated state-of-the-art performance in tasks such as audio-visual zero-shot classification, as shown by Li et al. \cite{li2023modality}. The aforementioned studies have demonstrated the powerful potential of SNNs in audio-visual joint learning. However, efficiently coupling SNNs with Transformers still faces following challenges:

\noindent\textbf{1) Time Steps}: Currently, most SNNs obtain the final output by averaging the output of each neuron with fixed time steps. These approaches not only overlook the importance of various layers in encoding temporal sequences but also cause significant fluctuations in SNN performance.

\noindent\textbf{2) Spiking Redundancy}: SNNs outputs exhibit redundancy with noise spikes present in both the temporal and spatial dimensions, which are highly correlated with spike firing frequency and neuron position. Finding a balance between spike neuron firing frequency and accuracy is crucial for reducing the redundancy of SNNs.

\noindent\textbf{3) Output Heterogeneity}: There is a significant difference in the output data distribution between SNNs and Transformers which are binary spike sequences and floating-point features, respectively. Efficiently integrating features from different data distributions is important to release the potential of SNNs.

To address the aforementioned challenges, we propose a new \textbf{S}piking \textbf{T}ucker \textbf{F}usion \textbf{T}ransformer (STFT) for audio-visual zero-shot learning in Fig. \ref{fig1}. Firstly, we introduce the time-step factors (TSF) which dynamically measure the significance of each time step in influencing the SNN's output. By efficiently utilizing the outputs from different time steps, these importance factors guide the synthesis of subsequent inference information. Additionally, we propose a global-local pooling (GLP) to combine the max and average pooling operation to guide the formation of the input membrane potential. The thresholds of the spiking neuron are adjust dynamically based on the semantic and temporal information cues. This helps reduce the generation of spike noise and improves the model's robustness. In terms of integrating the temporal information extracted by SNNs and the semantic information extracted by Transformers, a straightforward approach is to use a bilinear model for complete second-order interaction. However, this can lead to a significant increase in the number of parameters. We introduce a temporal-semantic Tucker fusion module to deal with this challenge. This module achieves multi-scale fusion of SNN and Transformer outputs at a very low cost while maintaining full second-order interactions.  We also demonstrate the qualitative comparison results with recent SOTA method MDFT in the bottom of Fig. \ref{fig1}. In sports classes with frequent changes in motion information, STFT demonstrate superiorities compared with MDFT due to the less spiking redundancy.

To sum up, our proposed SFTF aims to address the challenges of time steps, spiking redundancy and output heterogeneity in coupling SNNs and Transformer, enabling efficient fusion and interaction between temporal and semantic information. The main contributions of this paper are as follows:
\begin{itemize}
\item We propose a novel Spiking Tucker Fusion Transformer (STFT) for audio-visual zero-shot learning. STFT efficiently couples SNNs with Transformers, and combines the temporal and semantic information in different time steps to format the robust representations.

\item The temporal-semantic tucker fusion is proposed to achieve multi-scale fusion of SNN and Transformer outputs while maintaining full second-order interactions. This module effectively integrates the temporal and semantic information, providing a comprehensive representation for audio-visual data.

\item To reduce the spike noise, we adjust the thresholds of spiking neurons based on semantic and temporal information dynamically. The GLP is proposed to guide the formation of input membrane potentials based on their global and local characteristics.
\end{itemize}
The extensive experimental results prove that SFTF shows superiorities among state-of-the-art methods. The ablation study also demonstrates the effectiveness of each key component of our proposed model. 

\section{Related Work}
\subsection{Audio-Visual Zero-Shot Learning}
With the development of the deep learning \cite{Chen_2022_CVPR,Chen_2023_ICCV,tip1,tip2,tip3} in recent years, audio-visual zero-shot learning has gained significant attention due to its potential applications in various domains such as violence detection \cite{violence1}, aerial scene recognition \cite{aerial1}, speech recognition \cite{speech1,speech2} and video classification \cite{av2,av3,av4,av5,av6,av7,av8,av9,av11,av12}. MARBLE \cite{MARBLE} provides a comprehensive benchmark for evaluating AI in music understanding, addressing the need for deep music representations, large-scale datasets, and a universal community-driven standard. IcoCap \cite{IcoCAP} improves video captioning by using easily-learned image semantics to diversify video content, helping captioners focus on relevant information. Finsta \cite{Finsta} enhances video-language models by using fine-grained spatio-temporal alignment with scene graph structures, improving performance on various tasks without needing to retrain from scratch. Chen et al. \cite{MAML} propose Co-Meta Learning, which improves self-supervised speaker verification by leveraging complementary audio-visual information and updating network parameters using a disagreement strategy and meta learning. MAViL \cite{MAViL} uses three forms of self-supervision to learn audio-visual representations, achieving state-of-the-art performance in audio-video classification and enhancing both multimodal and unimodal tasks. SEEG \cite{SEEG} generates semantic-aware gestures by decoupling semantic-irrelevant information and leveraging semantic learning, outperforming other methods in semantic expressiveness and evaluations on various benchmarks. Hong et al. \cite{hong2023hyperbolic} incorporate a novel loss function that aligns video and audio features in the hyperbolic space, along with exploring the use of multiple adaptive curvatures for hyperbolic projections. Gowda et al. \cite{av10} discuss the issue of invalidation of the zero-shot setting in action recognition due to the overlap between classes in the pre-training and the evaluation datasets. They also highlight similar issues in few-shot action recognition and provide their splits for future evaluation in the field. Narayan et al. \cite{av1} incorporate a feedback loop from a semantic embedding decoder to refine the generated features iteratively. These synthesized features, along with their corresponding latent embeddings, are then transformed into discriminative features and utilized for classification, reducing ambiguities between categories. Wu et al. \cite{wuiccv19} propose the Dual Attention Matching (DAM) module, which spans longer video durations to better model high-level event information and captures local temporal details using a global cross-check mechanism. MA \cite{wu21cvpr} focus on improving weakly-supervised audio-visual video parsing by using cross-modal correspondence and contrastive learning to generate reliable event labels and address audio-visual asynchrony. Wu et al. \cite{wu2022switchable} propose the switchable LSTM framework to manage the generation and retrieval of nouns from external knowledge. Our proposed knowledge slots are primarily used for cross-modal fusion and semantic reasoning across different types of data (audio and visual). In contrast, the external knowledge in \cite{wu2022switchable} is specifically tailored for enhancing language models by incorporating external visual knowledge for novel object captioning. Yang et al. \cite{revised18} provides a comprehensive framework for multiple knowledge representations, which is crucial for understanding the integration of different modalities in ZSL. Yan et al. \cite{revised19} introduce a semantics-guided approach for zero-shot learning, which aligns closely with the temporal-semantic integration. Li et al. \cite{li2023modality,MDFT} first demonstrate the potential of SNN in audio-visual zero-shot learning. By efficiently extracting temporal information from different modalities using SNN, they achieved significant improvements in ZSL. However, due to the output heterogeneity between SNN and Transformer, their model's performance on seen classes tends to decline. Therefore, how to relieve this challenge is the key to release the potential of audio-visual ZSL.
\subsection{Spiking Neural Network}
Spiking Neural Networks (SNNs) are a biologically-inspired models that have time-evolving states \cite{snn1,snn2}. Unlike traditional neural networks that use continuous-valued activations, SNNs communicate through discrete spikes, which are analogous to action potentials in biological neurons. Each neuron in SNN receives input from neighboring neurons and generates a spike when the combined signals surpass a certain threshold. The precise timing of these spikes is crucial as it corresponds to the timing of action potentials in biological neurons. Spikes are transmitted between neurons through synapses, which have weights that determine the strength of the connections. The correlation between pre- and post-synaptic spikes is commonly used to train an SNN by adjusting the synaptic weights. In SNNs, information is encoded in the precise timing of spikes, allowing them to capture the temporal dynamics of data. The timing of spikes carries important information about the input, and the interactions between neurons are determined by the arrival times of these spikes. Currently, a significant number of researchers have been studying the intrinsic nature of SNNs, including attention mechanisms \cite{attsnn1,attsnn2,attsnn3}, deep SNNs \cite{deepsnn1,deepsnn2,deepsnn3,deepsnn4}, and simulations of biological visual pathways \cite{pipsnn}. In addition to these investigations, SNNs have found wide-ranging applications in various fields, such as image classification \cite{imagesnn1,imagesnn2}, speech recognition \cite{audiosnn1,audiosnn2}, object detection \cite{objectsnn1,objectsnn2}, and multimedia learning \cite{multisnn1,li2023modality}.

\begin{table}
\centering

\caption{Key Notations and Descriptions}
\label{tabkey}
\renewcommand{\arraystretch}{1.2}
\setlength{\tabcolsep}{3pt}
\begin{tabular}{|c|p{5.5cm}|}

\hline
\textbf{Notation} & \textbf{Description} \\ \hline
(G)ZSL & (Generalized) Zero-Shot Learning \\ \hline
STFT & Spiking Tucker Fusion Transformer \\ \hline
TSF & Time-Step Factor \\ \hline
GLP & Global-Local Pooling \\ \hline
HM & Harmonic Mean \\ \hline
SNN & Spiking Neural Network \\ \hline
$\boldsymbol{a}_i \in \mathbb{R}^{a_{\text{in}} \times h_{\text{emb}}}$ & Audio feature vector for sample $i$ \\ \hline
$\boldsymbol{v}_i \in \mathbb{R}^{v_{\text{in}} \times h_{\text{emb}}}$ & Visual feature vector for sample $i$ \\ \hline
$\boldsymbol{t}_i \in \mathbb{R}^{h_{\text{emb}}}$ & Textual embedding for sample $i$ \\ \hline
$E_a, E_v$ & Audio encoder, Visual encoder \\ \hline
$\boldsymbol{A}_t \in \mathbb{R}^{a_{\text{in}} \times h_{\text{emb}}}$ & Output of the audio encoder \\ \hline
$\boldsymbol{V}_t \in \mathbb{R}^{v_{\text{in}} \times h_{\text{emb}}}$ & Output of the visual encoder \\ \hline
$\boldsymbol{R}_a, \boldsymbol{R}_v \in \mathbb{R}^{h_{\text{emb}} \times h_{\text{emb}}}$ & Audio and visual latent semantic representations \\ \hline
$\boldsymbol{S}_c, \boldsymbol{\mathcal{G}}$ & Combined spiking output and Tucker core tensor \\ \hline
$\boldsymbol{K}_t \in \mathbb{R}^{h_{\text{emb}} \times h_{\text{emb}}}$ & Latent knowledge slots \\ \hline
$\boldsymbol{P}_a, \boldsymbol{P}_v$ & Projections of audio and visual features\\ \hline
$\boldsymbol{\mathcal{T}}_a \in \mathbb{R}^{d_{as} \times d_{at} \times K_a}$ & Audio tensors in Tucker decomposition \\ \hline
$\boldsymbol{\mathcal{T}}_v \in \mathbb{R}^{d_{vs} \times d_{vt} \times K_v}$ & Visual tensors in Tucker decomposition \\ \hline
$\boldsymbol{U}^{(s)} \in \mathbb{R}^{d_s \times n_a}$ & Factor matrices for spatial dimensions \\ \hline
$\boldsymbol{U}^{(t)} \in \mathbb{R}^{d_t \times n_v}$ & Factor matrices for temporal dimensions \\ \hline
$\boldsymbol{U}^{(k)} \in \mathbb{R}^{K \times n_k}$ & Factor matrices for latent dimensions \\ \hline
\end{tabular}
\end{table}

\section{Our Method}
The architecture of the STFT is illustrated in Fig. \ref{fig2} which consists of four primary components: spatial-temporal SNN, latent semantic reasoning module, temporal-semantic tucker fusion and joint reasoning module.

In the training phase, the seen classes training set, denoted as $\boldsymbol{\mathcal{X}}=(\boldsymbol{a}_{i}^{x},\boldsymbol{v}_{i}^{x},\boldsymbol{t}_{i}^{x})$, consists of $N$ samples. Here, $\boldsymbol{a}_{i}^{x}$ represents the audio feature, $\boldsymbol{v}_{i}^{x}$ represents the visual feature, and $\boldsymbol{t}_{i}^{x}$ represents the textual labeled embedding of the corresponding ground-truth class. The goal of STFT is to learn a projection function $f(\boldsymbol{a}_{i}^{x},\boldsymbol{v}_{i}^{x})\mapsto \boldsymbol{g}_{j}^{y}$, where $\boldsymbol{g}_{j}^{x}$ represents the class-level textual embedding for class $j$. This projection function is learned using the seen classes training set $\boldsymbol{\mathcal{X}}$. In the testing phase, the unseen testing set $\boldsymbol{\mathcal{Y}}=(\boldsymbol{a}_{i}^{y},\boldsymbol{v}_{i}^{y},\boldsymbol{t}_{i}^{y})$ is also projected using the function $f(\boldsymbol{a}_{i}^{y},\boldsymbol{v}_{i}^{y})\mapsto \boldsymbol{g}_{j}^{y}$. Overall, the STFT aims to learn a projection function that maps audio and visual features to class-level textual embeddings, allowing for the projection of unseen testing samples into the same embedding space. Table \ref{tabkey} demonstrated the notations and descriptions in detail.

\subsection{Latent Semantic Information Modeling}
\subsubsection{Audio and visual encoder} 
We employ the pre-trained SeLaVi model \cite{selavi} to accurately and effectively extract audio and visual features, as described in \cite{avca}. In order to further investigate the connections between contextual semantic information, we introduce audio and visual encoders, denoted as $E_{a}$ and $E_{v}$. The outputs of the audio and visual encoder can be written as: $\boldsymbol{\mathcal{A}}^{t}=E_{a}(\boldsymbol{\mathcal{X}}_{a})$ and $ \boldsymbol{\mathcal{V}}^{t}=E_{v}(\boldsymbol{\mathcal{X}}_{v})$, where $\boldsymbol{\mathcal{A}}^{t} \in \mathbb{R}^{a_{in}\times h_{emb}}$ and $\boldsymbol{\mathcal{V}}^{t} \in \mathbb{R}^{v_{in}\times h_{emb}}$. Each encoder for different modalities consists of two linear layers, namely $f_{1}^{s}$ and $f_{2}^{s}$ for $s \in (\boldsymbol{a}_{t},\boldsymbol{v}_{t})$. $f_{1}^{s}$: $\mathbb{R}^{s_{in}\times h_{in}} \rightarrow \mathbb{R}^{s_{in}\times h_{hid}}$ and $f_{2}^{m}$: $\mathbb{R}^{s_{in}\times h_{hid}} \rightarrow \mathbb{R}^{s_{in}\times h_{emb}}$. Each linear layer is followed by batch normalization, ReLU activation function, and dropout with a dropout rate of $d_{enc}$.

\subsubsection{Latent semantic reasoning module}
To better explore the potential relationships between semantic features within different modalities, we introduce the latent semantic reasoning module. We have observed that semantic features in audio and visual have correlations across different temporal dimensions. Therefore, we propose the Latent Knowledge Combiner (LKC) to dynamically update the latent semantic features of tow modalities, optimizing the feature representations of each modalities. The LKC assists in exploring and aligning latent cross-modal latent relationships, enabling to extract more robust multimodal feature representations.

\begin{figure*}
    \centering
	\includegraphics[scale=0.56]{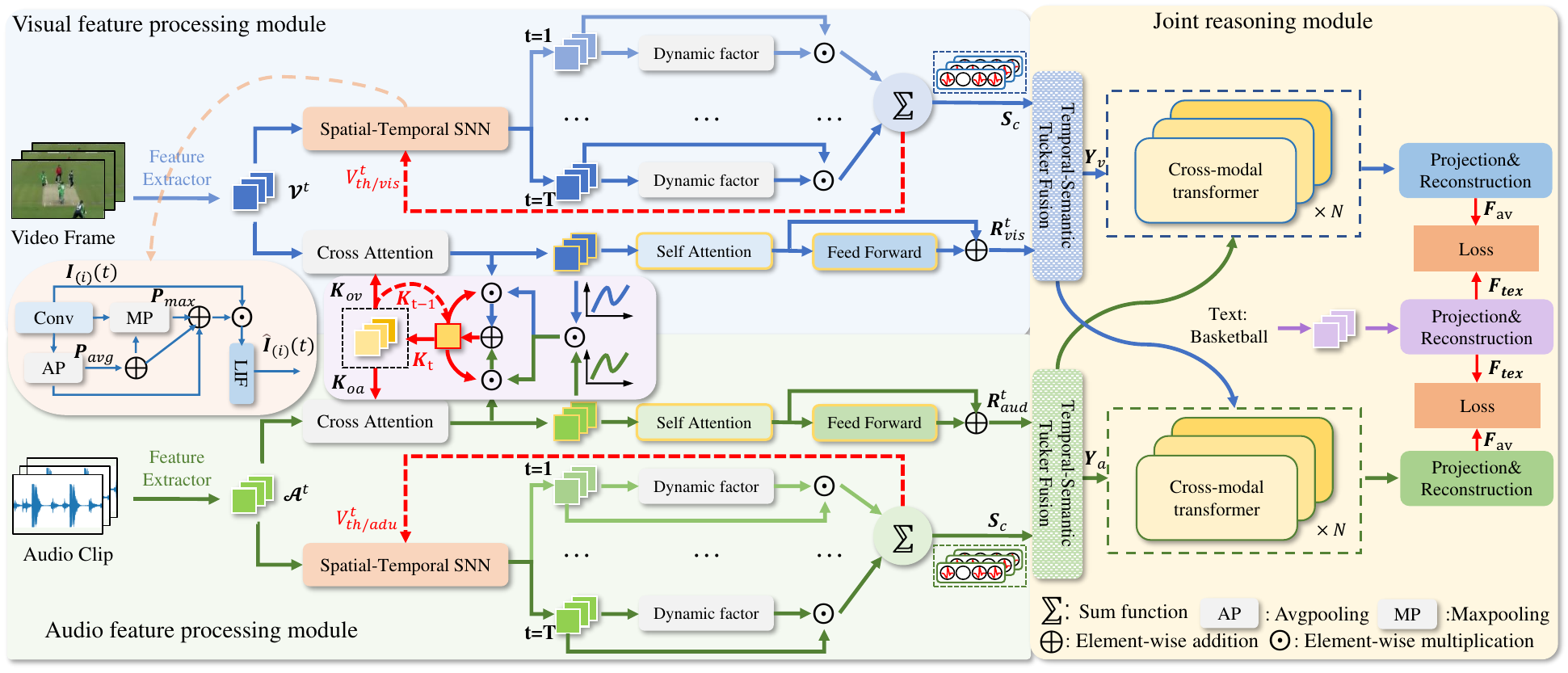}
	\caption{The overall architecture of STFT. The SNN thresholds are adjusted dynamically based on the semantic and temporal information cues. The spatial-temporal SNN using the GLP to refine the input features, combining the time-step factor to optimize the final output. The latent knowledge slots $\boldsymbol{K}_{t}$ could explore and align the latent semantic relationships of different modalities. The cross-modal transformer in joint reasoning module are shared weight.}
	\label{fig2}
\end{figure*}

The LKC captures a set of latent knowledge slots denoted as $\boldsymbol{K} = \{\boldsymbol{K}_{1},\boldsymbol{K}_{2},\ldots,\boldsymbol{K}_{n}\}$. These knowledge slots represent the latent semantic features that exist between two modalities. The illustration of LKC is shown in purple area of Fig. \ref{fig2}. The LKC could compute the importance of each latent knowledge slot based on the input vectors and effectively combines them together. Mathematically, this process can be expressed as follows:
\begin{equation}
\boldsymbol{K}_{oa} =\sum_{i=1}^{k} \phi(\boldsymbol{K}_{i}\boldsymbol{\mathcal{A}}_{t})\boldsymbol{\mathcal{A}}_{t},
  \boldsymbol{K}_{ov} =\sum_{i=1}^{k} \phi(\boldsymbol{K}_{i}\boldsymbol{\mathcal{V}}_{t})\boldsymbol{\mathcal{V}}_{t},
\end{equation}
where $\phi(x) = 1/(1+e^{-x})$. We proposed gate function to selectively remain the fusion features, which is defined as:
\begin{equation}
\begin{aligned}
    \boldsymbol{P}_{a} &= \mathrm{ReLU}(\boldsymbol{W}_{oa}\boldsymbol{K}_{oa}+b_{oa}),\\
    \boldsymbol{P}_{v} &= \mathrm{ReLU}(\boldsymbol{W}_{ov}\boldsymbol{K}_{ov}+b_{ov}),
\end{aligned}
\end{equation}
where $\boldsymbol{W}_{oa}\in \mathbb{R}^{h_{emb}\times h_{emb}}$ and $\boldsymbol{W}_{ov}\in \mathbb{R}^{h_{emb}\times h_{emb}}$ are learnable weight metrics and $b_{oa}$ and $b_{ov}$ are the bias items. The latent knowledge is update sequentially to connect the features of different modalities with previous latent knowledge slots $\boldsymbol{K}_{t-1}\in \mathbb{R}^{h_{emb}\times h_{emb}}$. The latent knowledge slots are updated as follows:
\begin{equation}
    \boldsymbol{K}_{t} = \alpha(\boldsymbol{P}_{a}\boldsymbol{K}_{oa}+\boldsymbol{P}_{v}\boldsymbol{K}_{ov}) + (1-\alpha) \boldsymbol{K}_{t-1}.
\end{equation}
where $\alpha$ is the learnable item to adjust the formation of the latent knowledge dynamically. The self-attention function is employed to further infer the inherent relationship between the audio and visual features using the latent knowledge. Formally, the outputs of the latent semantic reasoning module $\boldsymbol{R}_{a}^{t} \in \mathbb{R} ^{h_{emb}\times h_{emb}}$ and $\boldsymbol{R}_{v}^{t} \in \mathbb{R} ^{h_{emb}\times h_{emb}}$ are defined as:
\begin{equation}
	\begin{aligned}
		\boldsymbol{R}_{a}^{t} &= \mathrm{MLP}(\mathrm{LN}(\mathrm{SA}(\boldsymbol{K}_{oa}^{t})))+\mathrm{SA}(\boldsymbol{K}_{oa}^{t}),\\
 \boldsymbol{R}_{v}^{t} &= \mathrm{MLP}(\mathrm{LN}(\mathrm{SA}(\boldsymbol{K}_{ov}^{t})))+\mathrm{SA}(\boldsymbol{K}_{ov}^{t}),
	\end{aligned}
\end{equation}
where $\mathrm{SA}(\cdot)$ represents the self attention function, $\mathrm{LN}(\cdot)$ represents the layer normalization and $\mathrm{MLP}(\cdot)$ represents the multi-layer perceptron. 

\subsection{Spatial-Temporal SNN}
Unlike existing SNNs in the field of multimodal learning, we have specifically designed our SNN for the audio-visual domain. Firstly, we recognize the importance of temporal encoding by leveraging the information from different time steps in the SNN. We propose a time step factor (TSF) to dynamically fuse the outputs from different time steps. Additionally, to reduce the spiking noise in the SNN output and enhance the model's robustness, we introduce a global-local pooling (GLP) to improve the overall performance and stability of the SNN by combining the max and average pooling operations. 

Our SNN network consists of three convolution SNN blocks, each comprising a convolution operation layer followed by a LIF-based layer \cite{lif}. The LIF model consists of an integration phase, where the neuron accumulates input currents, and a firing phase, where the neuron generates a spike and resets its membrane potential. Specifically, the dynamics of a LIF neuron can be described by the following equation:
\begin{equation}
\tau_{m}\frac{d\boldsymbol{V}(t)}{dt}=-\boldsymbol{V}(t)+R\boldsymbol{I}(t),
\end{equation}
where $\tau_{m}$ is the membrane time constant, $\boldsymbol{V}(t)$ is the membrane potential at time $t$, $R$ is the membrane resistance, and $\boldsymbol{I}(t)$ is the current input at time $t$. To compute the input of the $i$-th LIF neuron $\boldsymbol{I}_{i}(t)$, we calculate the convolution operation and batch normalization with the output of the previous layer $\boldsymbol{P}(t)$ as:
\begin{equation}
\boldsymbol{I}_{i}(t)=\mathrm{BN}(\mathrm{CONV}(\boldsymbol{W}_{p},\boldsymbol{P}(t))),
\end{equation}
where $\boldsymbol{W}_{p}$ represents the weight matrix, $\mathrm{BN}(\cdot)$ represents the batch normalization and $\mathrm{CONV}(\cdot)$ represents the convolution operation. When the membrane potential reaches a threshold value $v_{th}$ would generate a spike, and the membrane potential is reset to a reset potential $V_{rest}$. 

To optimize the distribution of input features before processing by the LIF neurons, we propose the GLP as shown in Fig. \ref{fig3}. The max pooling operation captures the global maximum value of the input features, which represents the overall variation in the input distribution. The average pooling operation calculates the average value of the input features, which highlights the importance of local salient regions. By combining the outputs of max and average pooling, the GLP module provides guidance for generating the input features based on the global and local characteristics. The overall process can be written as follows:
\begin{equation}
	\begin{aligned}
        \boldsymbol{P}_{all} &= \frac{1}{2}(\boldsymbol{P}_{max}+\boldsymbol{P}_{avg})+\beta \boldsymbol{P}_{max}+ (1-\beta)\boldsymbol{P}_{avg},\\
        \hat{\boldsymbol{I}_{i}(t)} &= \phi (\boldsymbol{P}_{all}\boldsymbol{I}_{i}(t)+\boldsymbol{I}_{i}(t))
	\end{aligned}
\end{equation}
where $\boldsymbol{P}_{max}$ and $\boldsymbol{P}_{avg}$ are corresponding to the max and average pooling features, and $\beta$ is the learnable items. Indeed, the relationship between the output of the SNN and the corresponding time steps is crucial. Effectively utilizing the outputs from different time steps can significantly influence the final performance. A common method is to assign equal weights to each time step and compute the average output across all time steps to obtain the final result. However, this method overlooks the diversity among different time steps. To deal with this, we propose the TSF which adjust the weights of SNN outputs in different time steps dynamically. By considering the importance of each time step, the SNN can effectively capture the temporal dynamics and encode relevant information at different time scales. The final output of the spatial-temporal SNN can be summarized by the following equations:
\begin{equation}
\boldsymbol{S}_{c} = \sum_{T}^{i=1}max(\frac{e^{\boldsymbol{I}_{i}(t)} }{ {\textstyle \sum_{j=1}^{T} e^{\boldsymbol{I}_{i}(j)}} } ) \boldsymbol{I}_{i}(t),
\end{equation}
where $max(\cdot)$ returns the largest item of the input and $c \in (a_{t},v_{t})$. This dynamic adjustment of weights allows the SNN to adaptively emphasize the contributions of different time steps based on their significance, leading to obtain more fine-grained representation of temporal sequences. 

To reduce the spiking noise, we dynamically adjusting the threshold of the LIF neurons based on the current output of the SNN and the pooling matrix. Specifically, we use the entropy of the current SNN output to represent the amount of information contained in the features. If the information content is rich, it indicates a more informative scene, and we need to increase the threshold to suppress spike noise. The threshold adjustment for the audio and visual modalities of the SNN can be expressed as follows:
\begin{equation}
	\begin{aligned}
        V_{th/aud}^{t} &= (\phi(\boldsymbol{P}_{all})+\mathcal{N}(\boldsymbol{S}_{a}^{t})log(\mathcal{N}(\boldsymbol{S}_{a})))V_{th/aud}^{t-1},\\
        V_{th/vis}^{t} &= (\phi(\boldsymbol{P}_{all})+\mathcal{N}(\boldsymbol{S}_{v}^{t})log(\mathcal{N}(\boldsymbol{S}_{v})))V_{th/vis}^{t-1},
	\end{aligned}
\end{equation}
where $\mathcal{N}(\cdot)$ represents the normalization operation.

\subsection{Temporal-Semantic Tucker Fusion}
In this paper, we utilize both the spatial-temporal SNN and LSR module to extract temporal and semantic information, respectively. However, these two types of network outputs have significantly different data distributions: one is a binary sequence, while the other is a floating-point feature. It poses a challenge to effectively fuse these outputs while preserving the complex and high-level interactions. A powerful solution for feature fusion that has been recently proposed is bilinear modeling, which can encode fully parameterized bilinear interactions. First, the semantic and temporal features in each modality need to be projected into embedding vectors as $\boldsymbol{R}_{a} \in \mathbb{R}^{d_{as}}$, $\boldsymbol{R}_{v} \in \mathbb{R}^{d_{vs}}$, $\boldsymbol{S}_{a} \in \mathbb{R}^{d_{at}}$ and $\boldsymbol{S}_{v} \in \mathbb{R}^{d_{vt}}$, respectively. The bilinear model in visual pipeline can be written as:
\begin{equation}
        \boldsymbol{Y}_{a}= \boldsymbol{\mathcal{T}_{a}} \times _{1}\boldsymbol{R}_{a}\times _{2}\boldsymbol{S}_{a},
        \boldsymbol{Y}_{v}= \boldsymbol{\mathcal{T}_{v}} \times _{1}\boldsymbol{R}_{v}\times _{2}\boldsymbol{S}_{v},
\end{equation}
where $\boldsymbol{\mathcal{T}_{a}} \in \mathbb{R} ^{d_{as}\times d_{at} \times K_{a}}$ and $\boldsymbol{\mathcal{T}_{v}} \in \mathbb{R} ^{d_{vs}\times d_{vt} \times K_{v}}$ represent the full tensors and $\times _{i}$ represents the $i$-mode product. However, the parameters in full tensor $\boldsymbol{\mathcal{T}_{c}}$ could be very large. Here, we propose the temporal-semantic tucker fusion to factorize the full tensor $\boldsymbol{\mathcal{T}_{c}}$ following \cite{mutan}. The decomposition of full tensor $\boldsymbol{\mathcal{T}}$ could be defined as :
\begin{equation}
        \boldsymbol{\mathcal{T}}: =\boldsymbol{\mathcal{G}}\times _{1}\boldsymbol{U}^{(s)}\times _{2}\boldsymbol{U}^{(t)}\times _{3}\boldsymbol{U}^{(k)},
\end{equation}
where $\boldsymbol{\mathcal{G}}$ is the core tensor, $\boldsymbol{U}^{(s)} \in \mathbb{R}^{d_{s}\times n_{a}}$, $\boldsymbol{U}^{(t)} \in \mathbb{R}^{d_{t}\times n_{v}}$ and $\boldsymbol{U}^{(k)} \in \mathbb{R}^{K\times n_{k}}$ are the factor matrices. We could utilize the tensor decomposition to factorize the full tensor $\boldsymbol{\mathcal{T}_{a}}$ and $\boldsymbol{\mathcal{T}_{v}}$, and rewrite the Eq. (10) as follows:
\begin{equation}
\begin{aligned}
        \boldsymbol{Y}_{a}:&= \boldsymbol{\mathcal{G}_{a}} \times _{1}(\boldsymbol{R}_{a}^{\top}\boldsymbol{U}^{t}_{a})\times _{2}(\boldsymbol{S}_{a}^{\top}\boldsymbol{U}^{s}_{a}))\times _{3}\boldsymbol{U}^{k}_{a},\\
        \boldsymbol{Y}_{v}:&= \boldsymbol{\mathcal{G}_{v}} \times _{1}(\boldsymbol{R}_{v}^{\top}\boldsymbol{U}^{t}_{v})\times _{2}(\boldsymbol{S}_{v}^{\top}\boldsymbol{U}^{s}_{v}))\times _{3}\boldsymbol{U}^{k}_{v}.
        \end{aligned}
\end{equation}

We can perform bilinear interaction to capture the complex relationships between the temporal and semantic information, and then project them into a lower-dimensional representation. This process is defined as as:
\begin{equation}
\begin{aligned}
        \boldsymbol{Y}_{a}&= \boldsymbol{\mathcal{G}_{a}} \times _{1}\widetilde{\boldsymbol{R}_{a}}\times _{2}\widetilde{\boldsymbol{S}_{a}},\\
        \boldsymbol{Y}_{v}&= \boldsymbol{\mathcal{G}_{v}} \times _{1}\widetilde{\boldsymbol{R}_{v}}\times _{2}\widetilde{\boldsymbol{S}_{v}},
\end{aligned}
\end{equation}
where $\widetilde{\boldsymbol{R}_{a}}=\boldsymbol{R}_{a}^{\top}\boldsymbol{U}^{s}_{a} \in \mathbb{R}^{n_{a}\times s_{a}}$, $\widetilde{\boldsymbol{R}_{v}}=\boldsymbol{R}_{v}^{\top}\boldsymbol{U}^{s}_{v} \in \mathbb{R}^{n_{v}\times s_{v}}$, $\widetilde{\boldsymbol{S}_{a}}=\boldsymbol{S}_{a}^{\top}\boldsymbol{U}^{t}_{a} \in \mathbb{R}^{n_{a}\times t_{a}}$ and $\widetilde{\boldsymbol{S}_{v}}=\boldsymbol{S}_{v}^{\top}\boldsymbol{U}^{t}_{v} \in \mathbb{R}^{n_{v}\times t_{v}}$.
\subsection{Joint Reasoning Module}
After integrating the temporal and semantic features from different modalities, we propose a cross-modal transformer to further reason about the implicit feature correspondences within each modality. We establish residual connections between the two modalities, to capture the complementary information between them. Layer normalization is applied to mitigate the impact of feature variations. The cross-modal transformer contains a stack of standard transformer layers to obtains a joint temporal-semantic representation. The cross-modal transformer block in two modalities are shared weight which can be summarized as follow:
 \begin{equation}
	\begin{aligned}
		\boldsymbol{Q}_{av} &= \mathrm{MHCA}(\boldsymbol{Y}_{a},\boldsymbol{Y}_{v}),\\
		\boldsymbol{Z}_{av} &= \mathrm{MLP}(\mathrm{LN}(\boldsymbol{Q}_{i})) + \boldsymbol{Q}_{av},
	\end{aligned}
\end{equation}
where $\mathrm{MHCA}(\cdot)$ represents the multi-head cross attention. The ultimate goal of our model is to predict the text category based on the audio-visual inputs. To project the joint audio-visual features into the same space as the text features, we construct the projection and reconstruction layers. The projection layer maps the audio-visual features to align with the text feature. the reconstruction layer helps to preserve the relevant information while discarding any noise or irrelevant details that may have been introduced during the projection. Both the projection and reconstruction layers have a similar structure, consisting of two linear layers  $f_{3}^{m}: \mathbb{R}^{s_{in}*h_{emb}} \rightarrow \mathbb{R}^{s_{in}*h_{hid}}$ and $f_{4}^{m}: \mathbb{R}^{s_{in}*h_{hid}} \rightarrow \mathbb{R}^{s_{in}*h_{out}}$, followed by dropout regularization with rate $d_{proj}$. The final audio-visual joint feature embeddings can be obtained as:
 \begin{equation}
	\begin{aligned}
        \boldsymbol{\mathcal{F}}_{av} &= \boldsymbol{Pro}_{av}(\boldsymbol{Z}_{av}),
	\end{aligned}
\end{equation}
where $\boldsymbol{Pro}_{av}(\cdot)$ is the projection function. The final textual labeled embedding $\boldsymbol{\mathcal{F}}_{tex}$ is obtained through the word projection layer $\boldsymbol{Pro}_{tex}$. The architecture of the $\boldsymbol{W}_{tex}$ is similar with $\boldsymbol{Pro}_{av}$ with dropout rate $d_{text}$.
\subsection{Training Strategy}
The STFT is tranined using a Nvidia V100S GPU. The audio and visual embeddings are extracted using pretrained SeLaVi \cite{selavi}. In STFT, we set $a_{in}=512$, $h_{emb}=512$, $h_{hid}=512$, $h_{out}=300$ and $h_{proj}=64$. In VGGSound, UCF, ActivitiNet datasets, the dropout rates are corresponding to $d_{enc} = (0.20,0.25,0.10)$, $d_{dec} = (0.25,0.20,0.15)$, and $d_{text} = (0.1,0.1,0.1)$, respectively. The cross-modal transformer is constructed with 8 heads, the dimension of each head is 64. We select Adam as training optimizer. STFT is trained 60 epochs with 0.0001 learning rate. To update parameters more effectively, STFT using the combination of triplet loss $\mathcal{L}_{t}$, projection loss $\mathcal{L}_{p}$ and reconstruction loss $\mathcal{L}_{r}$.

\subsubsection{Triplet loss.}
The triplet loss compares the distances between anchor samples, positive samples, and negative samples in the joint audio-visual embedding space. The triplet loss $\mathcal{L}_{t}$ can be written as:
\begin{equation}
\begin{split}
    \mathcal{L}_{t}=[\gamma +\boldsymbol{ \mathcal{F}}_{av}^{+}-\boldsymbol{ \mathcal{F}}_{tex}^{+}]_{+}
	+[\gamma +\boldsymbol{ \mathcal{F}}_{av}^{-}-\boldsymbol{ \mathcal{F}}_{tex}^{+}]_{+},
\end{split}
\end{equation}
where $\gamma$ represents the crucial margin parameter that defines the minimum separation between negative pairs of different modalities and truly matching audio-visual embeddings, $\boldsymbol{ \mathcal{F}}_{tex}$ represents the textual embeddings, $ [x]_{+}\equiv \max(x,0)$, and $\boldsymbol{\mathcal{F}}_{av}^{+}$ and $\boldsymbol{\mathcal{F}}_{av}^{-}$ correspond to positive and negative examples respectively.

\subsubsection{Projection loss.}
The projection loss reduce the distance between the output joint embeddings from the projection layer and the corresponding textual labeled embeddings, which can be written as:
\begin{equation}
\mathcal{L}_{p} = \frac{1}{n} \sum_{i=1}^{n}(\boldsymbol{ \mathcal{F}}_{av}-\boldsymbol{ \mathcal{F}}_{tex}),
\end{equation}
where $n$ is the number of samples.

\subsubsection{Reconstruction loss.}
The reconstruction loss is proposed to ensure the original data distribution is maintained when projecting audio-visual features to the shared embedding space. The reconstruction loss $\mathcal{L}_{r}$ can be written as:
\begin{equation}
\mathcal{L}_{r} = \frac{1}{n} \sum_{i=1}^{n}(\boldsymbol{ \mathcal{F}}_{av}^{rec}-\boldsymbol{ \mathcal{F}}_{tex}),
\end{equation}
where $\boldsymbol{\mathcal{O}}_{av}^{rec}$ is the output of the reconstruction layer. The total loss is formulated as $\mathcal{L}_{all} = 0.5*\mathcal{L}_{t}+0.5*(\mathcal{L}_{p}+\mathcal{L}_{r})$.
\section{Experiment}
In this paper, we evaluate our proposed model in both ZSL and GZSL settings. Following \cite{avca}, we utilize the mean class accuracy to measure the effectiveness of the models in classification tasks. For the ZSL evaluation, we specifically focus on analyzing the performance of the models on test samples from the subset of unseen test classes. In the GZSL evaluation, we evaluate the models on the entire test set, which includes both seen (S) and unseen (U) classes. This comprehensive evaluation enables us to calculate the harmonic mean, which is given by the formula: $\mathrm{HM}=\frac{2\mathrm{US}}{\mathrm{U}+\mathrm{S}}$. The harmonic mean provides a balanced measure of the model's overall performance in the GZSL scenario.

\begin{table*}
	\centering
  \renewcommand\arraystretch{1.5}
	\begin{threeparttable}

		\caption{The performance of our STFT and state-of-the-art baselines for audio-visual (G)ZSL on three benchmark datasets.}
		\label{TAB1}
		\setlength{\tabcolsep}{6pt}{
\begin{tabular}{cccccclcccclcccc}
\hline \hline
\multirow{2}{*}{Type} &
  \multirow{2}{*}{Model} &
  \multicolumn{4}{c}{VGGSound-GZSL} &
   &
  \multicolumn{4}{c}{UCF-GZSL} &
   &
  \multicolumn{4}{c}{ActivityNet-GZSL} \\ \cline{3-6} \cline{8-11} \cline{13-16} 
 &
   &
  S &
  U &
  \textit{HM $\uparrow$} &
  \textit{ZSL $\uparrow$} &
   &
  S &
  U &
  \textit{HM $\uparrow$} &
  \textit{ZSL $\uparrow$} &
   &
  S &
  U &
  \textit{HM $\uparrow$} &
  \textit{ZSL $\uparrow$} \\ \hline
\multirow{2}{*}{ZSL} &
 SJE \cite{SJE} &
  48.33 &
  1.10 &
  2.15 &
  4.06 &
   &
  63.10 &
  16.77 &
  26.50 &
  18.93 &
   &
  4.61 &
  7.04 &
  5.57 &
  7.08 \\
 &
 DEVISE \cite{devise} &
  36.22 &
  1.07 &
  2.08 &
  5.59 &
   &
  55.59 &
  14.94 &
  23.56 &
  16.09 &
   &
  3.45 &
  8.53 &
  4.91 &
  8.53 \\
 &
  APN \cite{APN} &
  7.48 &
  3.88 &
  5.11 &
  4.49 &
   &
  28.46 &
  16.16 &
  20.61 &
  16.44 &
   &
  9.84 &
  5.76 &
  7.27 &
  6.34 \\
 &
  VAEGAN \cite{VAEGAN} &
  12.77 &
  0.95 &
  1.77 &
  1.91 &
   &
  17.29 &
  8.47 &
  11.37 &
  11.11 &
   &
  4.36 &
  2.14 &
  2.87 &
  2.40 \\ \hline
\multirow{5}{*}{\begin{tabular}[c]{@{}c@{}}Audio-visual\\ ZSL\end{tabular}} &
  CJME \cite{cjme} &
  8.69 &
  4.78 &
  6.17 &
  5.16 &
   &
  26.04 &
  8.21 &
  12.48 &
  8.29 &
   &
  5.55 &
  4.75 &
  5.12 &
  5.84 \\
 &
  AVGZSLNet \cite{avgzslnet} &
  18.05 &
  3.48 &
  5.83 &
  5.28 &
   &
  52.52 &
  10.90 &
  18.05 &
  13.65 &
   &
  8.93 &
  5.04 &
  6.44 &
  5.40 \\
 &
  AVCA \cite{avca} &
  14.90 &
  4.00 &
  6.31 &
  6.00 &
   &
  51.53 &
  18.43 &
  27.15 &
  20.01 &
   &
  24.86 &
  8.02 &
  12.13 &
  9.13 \\
 &
  TCaF \cite{tcaf} &
  9.64 &
  5.91 &
  7.33 &
  6.06 &
   &
  58.60 &
  21.74 &
  31.72 &
  24.81 &
   &
  18.70 &
  7.50 &
  10.71 &
  7.91 \\ 
 &
  AVMST \cite{li2023modality}&
  14.14 &
  5.28 &
  7.68 &
  6.61 &
   &
  44.08 &
  22.63 &
  29.91 &
  28.19 &
   &
  17.75 &
  9.90 &
  12.71 &
  10.37 \\& 
Hyper$^{\mathrm{alignment}}$ \cite{hong2023hyperbolic} &
  13.22 &
  5.01 &
  7.27 &
  6.14 &
   &
  57.28 &
  17.83 &
  27.19 &
  19.02 &
   &
  23.50 &
  8.47 &
  12.46 &
  9.83
 \\
 &
Hyper$^{\mathrm{single}}$ \cite{hong2023hyperbolic}&
  9.79 &
  6.23 &
  7.62 &
  6.46 &
   &
  52.67 &
  19.04 &
  27.97 &
  22.09 &
   &
  23.60 &
  10.13 &
  14.18 &
  10.80 
 \\
 &
Hyper$^{\mathrm{multiple}}$ \cite{hong2023hyperbolic} &
  15.02 &
  6.75 &
  9.32 &
  7.97 &
   &
  63.08 &
  19.10 &
  29.32 &
  22.24 &
   &
  23.38 &
  8.67 &
  12.65 &
  9.50 
 \\ &
    MDFT \cite{MDFT} &
  16.14 &
  5.97 &
  8.72 &
  7.13 &
   &
  48.79 &
  23.11 &
  31.36 &
  \textbf{31.53} &
   &
  18.32 &
  10.55 &
  13.39 &
  12.55 \\ \hline \rowcolor{lightgray!30}
& STFT (ours) &
  19.22 &
  6.81 &
  \textbf{10.06}&
  \textbf{8.24} &
   &
  56.47 &
  22.89 &
\textbf{32.58}&
  29.72 &
   &
  22.34 &
  11.73 &
  \textbf{15.38} &
  \textbf{12.91} \\
\hline \hline
\end{tabular}

}
\end{threeparttable}
\end{table*}

\subsection{Dataset Statistics}
In this study, we conducted experiments and evaluated the proposed models using three benchmark datasets: ActivityNet, VGGSound, and UCF101. These datasets were chosen to provide a diverse range of audio-visual data and cover various domains, enabling a comprehensive evaluation of the proposed models' performance. The statistics of these datasets are as follows: 1). \textbf{ActivityNet} contains a wide variety of human activities along with the corresponding videos. The dataset consists of approximately 200 different activity classes and more than 20,000 videos with an average duration of about 2 minutes per video. 2). \textbf{UCF101} dataset consists of more than 13,000 videos collected from YouTube, with an average duration of around 7 seconds per video. The videos cover a wide range of human actions in various contexts and provide a challenging dataset for action recognition algorithms. 3). \textbf{VGGSound} dataset consists of more than 200 different classes and includes thousands of audio clips obtained from online sources.

\subsection{Results Comparison}
In Table \ref{TAB1}, we demonstrate the superiority of proposed STFT compared to state-of-the-art (SOTA) methods. On the VGGSound dataset, STFT achieves significant improvements over TCaF with a 37.2\% increase in HM and a 35.9\% increase in ZSL scores. On the UCF101 dataset, STFT achieves an HM of 32.58 and a ZSL score of 29.72. Compared to the best current method MDFT \cite{MDFT}, STFT achieves a 3.9\% improvement in HM but experiences a slight decrease in ZSL. It's worth noting that both our STFT and MDFT models employ SNN as temporal encoders. However, due to the different output data distributions between SNN and the Transformer used in MDFT, MDFT's performance on Seen Classes is not satisfactory. To address this issue, we propose the temporal-semantic tucker fusion, which achieves a 15.7\% improvement on Seen Classes. On the ActivitiNet dataset, STFT obtains an HM score of 15.38 and a ZSL score of 12.91, surpassing AVCA's HM score of 12.13 and ZSL score of 9.13. Compared to AVMST, STFT achieves a 21\% improvement in HM and a 24.5\% improvement in ZSL. Overall, our STFT model demonstrates superior performance compared to existing methods on various evaluation metrics across the three datasets.

While the proposed method shows significant improvements on the VGGSound-GZSL dataset compared to others, it is important to note that the MDFT method requires data enhancement to convert RGB images to events, inherently introducing additional computational complexity. MDFT utilizes an Event Generative Model (EGM) to convert RGB images into event streams, eliminating background scene bias and capturing motion information. However, this conversion adds computational overhead, requiring the processing of high-resolution image data to generate events and the use of Spiking Neural Networks (SNNs) to handle the sparse event data efficiently. In contrast, our proposed method avoids this complexity by directly modeling the audio-visual data without converting RGB images to events. This design choice allows our method to maintain competitive performance across different datasets while being more computationally efficient.

Moreover, we observed a slight decline in ZSL performance on the UCF101 dataset. This decline can be attributed to the fixed rank constraint used in the semantic-temporal Tucker fusion module, which may not fully capture the complex temporal dynamics of the UCF101 dataset. To address this issue, we suggest dynamically adjusting the rank constraint based on the singular values of the input data in the future. Additionally, significant variations in activity patterns within the UCF101 dataset may introduce redundancy at higher time steps, negatively impacting ZSL performance. We suggest reducing redundancy through an optimized temporal encoding process and exploring different configurations of the time-step factor to enhance temporal feature integration.

\begin{table}
	\centering
        \renewcommand\arraystretch{1.5}
	\begin{threeparttable}
		\caption{Ablation study of different loss items.}
		\label{TAB2}
		\setlength{\tabcolsep}{11pt}{
\begin{tabular}{ccccc}
\hline \hline
\multirow{2}{*}{Loss}            & \multicolumn{4}{c}{UCF-GZSL}                             \\ \cline{2-5} 
                                  & S     & U              & \textit{HM $\uparrow$}    & \textit{ZSL $\uparrow$}   \\ \hline
W/o $\mathcal{L}_{p}$+$\mathcal{L}_{r}$  & 48.76 & 17.21          & 25.44          & 19.46          \\
W/o $\mathcal{L}_{p}$  & 53.14 & 18.21          &  27.13         & 23.14           \\
W/o $\mathcal{L}_{r}$  & 51.47 & 19.33          &  28.10         & 23.81           \\ \hline \rowcolor{lightgray!30}
STFT                              & 56.47 & 22.89 & \textbf{32.58} & \textbf{29.72} \\
\hline \hline
\end{tabular}
		}
	\end{threeparttable}
\end{table}
\begin{table}
	\centering
        \renewcommand\arraystretch{1.5}
	\begin{threeparttable}
		\caption{Ablation study of different model components.}
		\label{TAB3}
		\setlength{\tabcolsep}{13pt}{
\begin{tabular}{ccccc}
\hline \hline
\multirow{2}{*}{Components}            & \multicolumn{4}{c}{UCF-GZSL}                             \\ \cline{2-5} 
                                  & S     & U              & \textit{HM $\uparrow$}    & \textit{ZSL $\uparrow$}   \\ \hline
W/o GLP & 52.88 & 18.72          &   27.65        & 24.79 \\
W/o TSF  & 52.14 & 19.44          &  28.32         & 25.52 \\
W/o DTH & 53.79 & 21.72           &  30.94         & 27.41 \\
W/o LKC  & 49.13 & 22.67            &  31.02         & 28.96          \\ \hline \rowcolor{lightgray!30}
STFT                              & 56.47 & 22.89 & \textbf{32.58} & \textbf{29.72} \\ \hline \hline
\end{tabular}
		}
	\end{threeparttable}
\end{table}

\begin{figure*}
  \centering
  \subfloat[HM on different time step.]{
    \includegraphics[width=6cm]{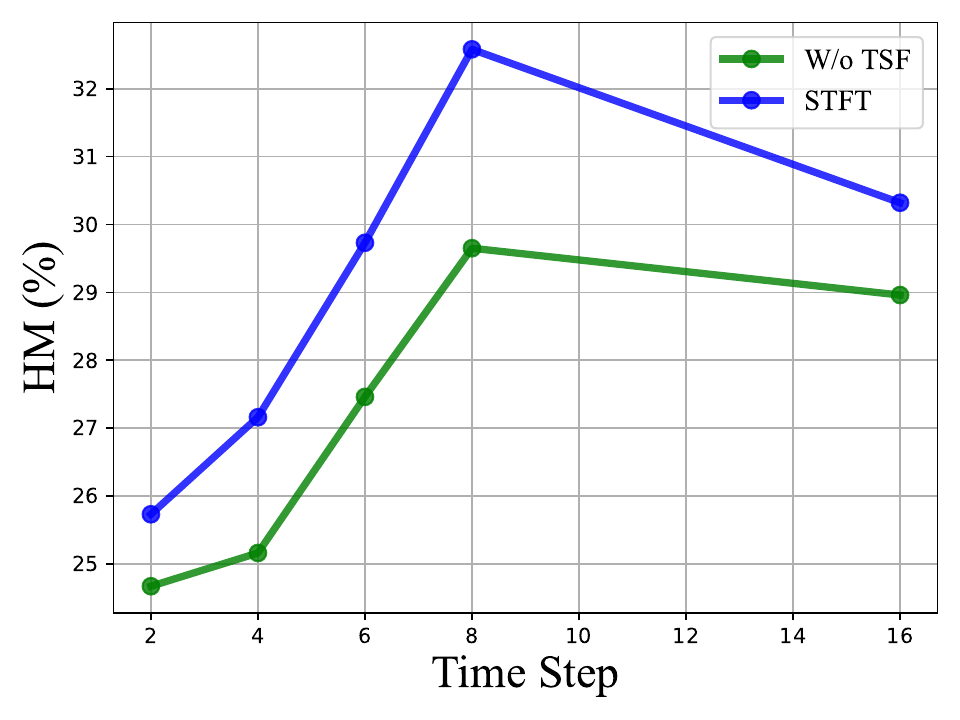}}  
  \subfloat[HM on different rank constraint.]
  {
    \includegraphics[width=6cm]{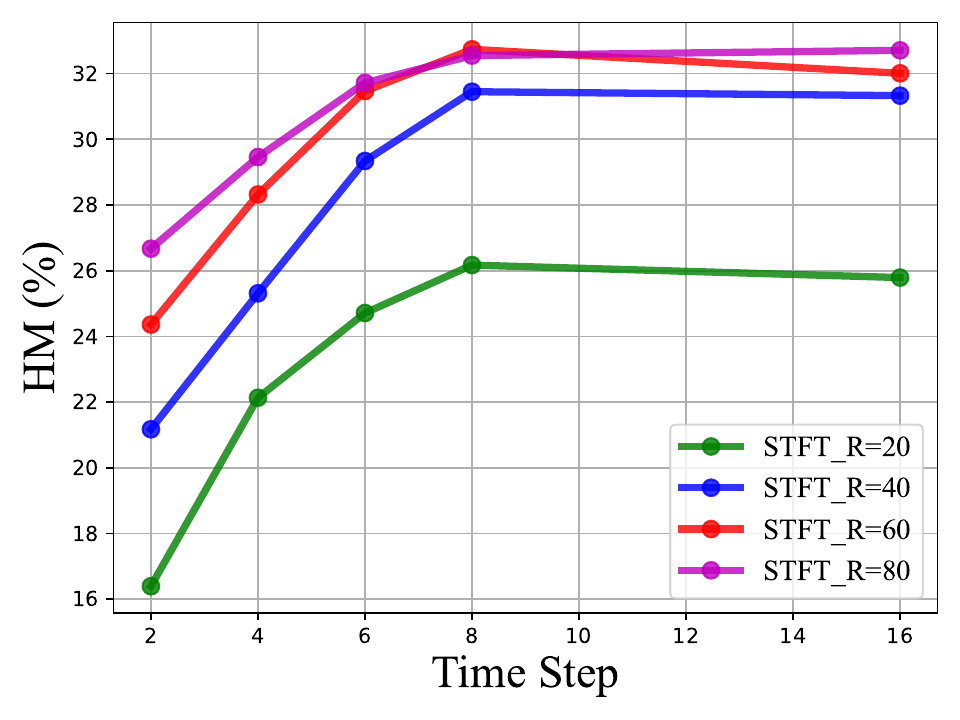}
  }
    \subfloat[HM on different fix threshold.]
  {
    \includegraphics[width=6cm]{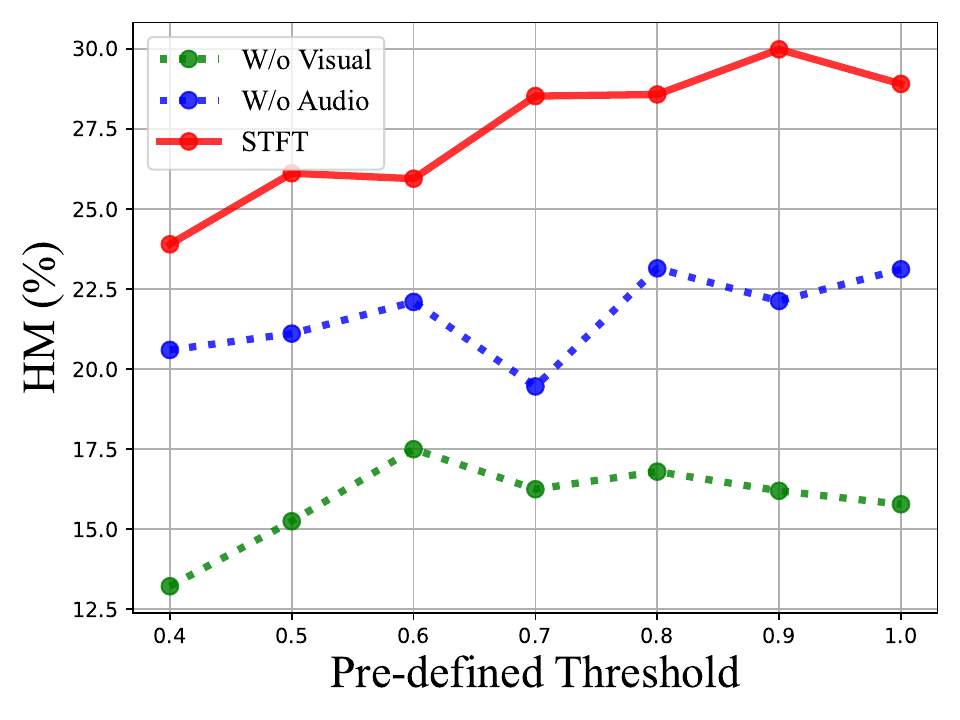}
  }

      \subfloat[ZSL on different time step.]
  {
    \includegraphics[width=6cm]{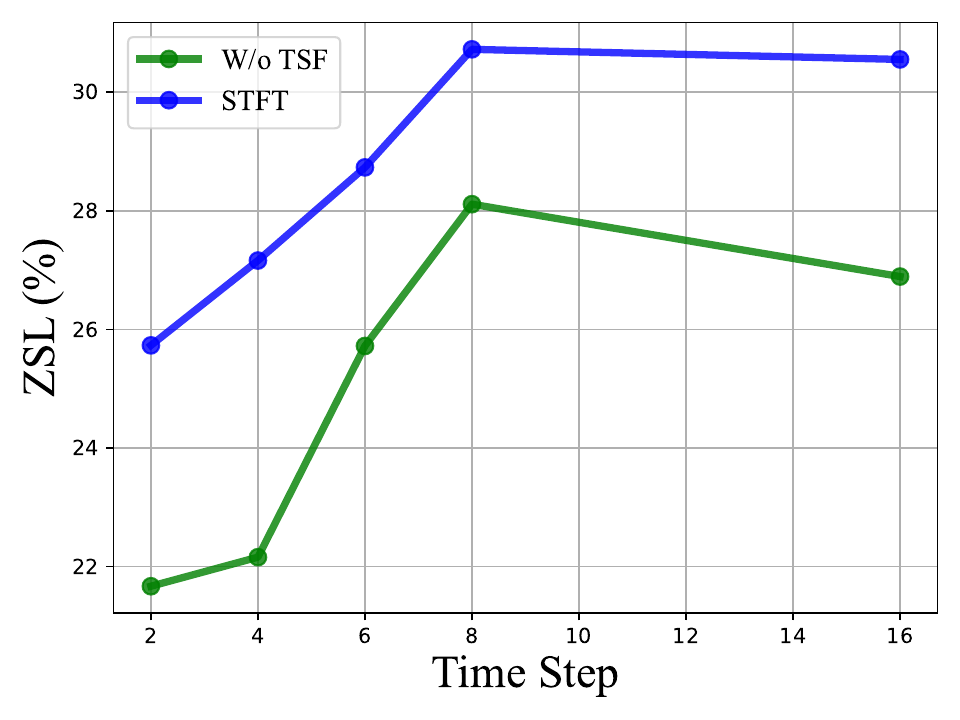}
  }
  \subfloat[ZSL on different rank constraint.]{
    \includegraphics[width=6cm]{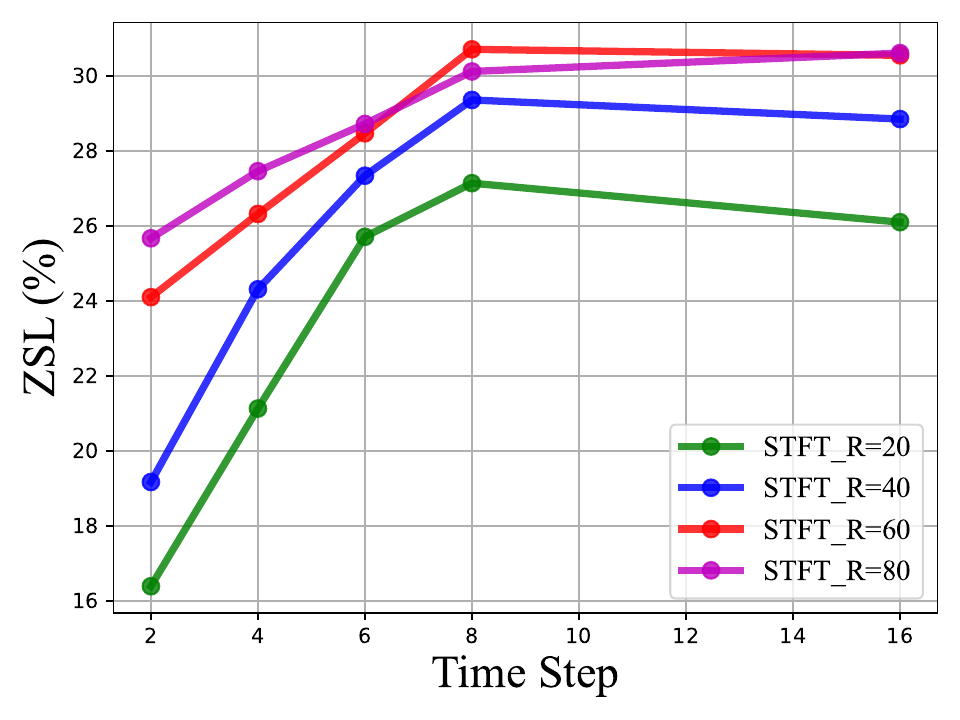}
  }
  \subfloat[ZSL on different fix threshold.]
  {
    \includegraphics[width=6cm]{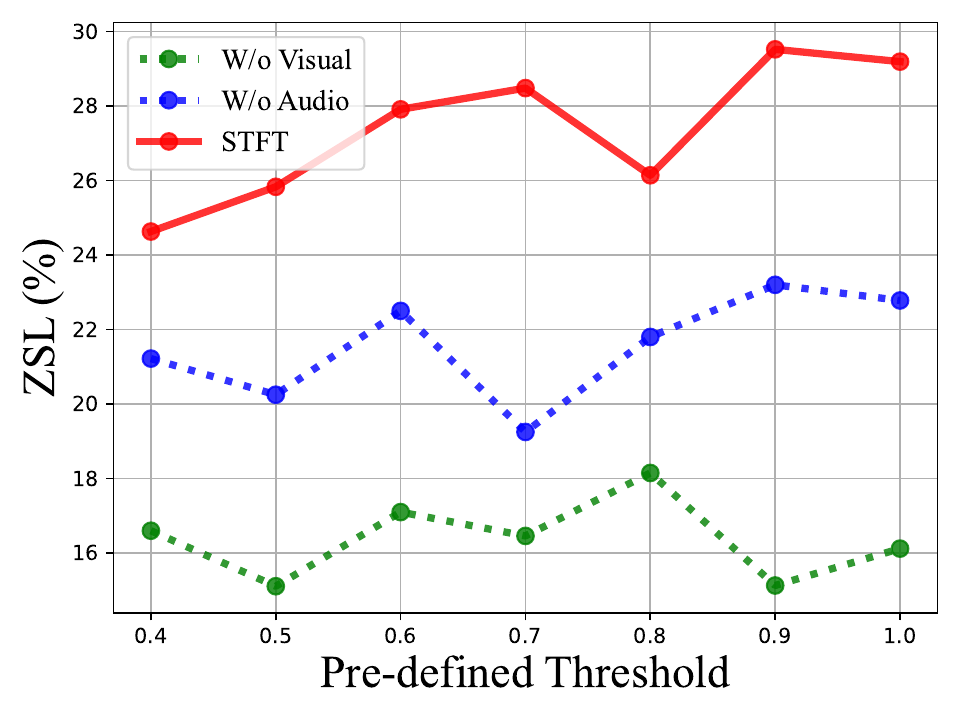}
  }
  \caption{The ablation study of the impact of different time step, rank constraint and fix thresholds to HM and ZSL performance on UCF101 dataset.}
  \label{fig3}
\end{figure*}

\subsection{Ablation Study}
\subsubsection{The effectiveness of different model components} 
To demonstrate the effectiveness of each component in our model, we conducted extensive experiments on the UCF dataset, as shown in Table \ref{TAB2}. The models without the Latent Knowledge Combiner, Global-Local Pooling module, Time-Step Factor, and Dynamic Threshold Adjustment module are denoted as ``W/o LKC," ``W/o GLP," ``W/o TSF," and ``W/o DTH," respectively. Among these components, the GLP module in the SNN has the most significant impact on the overall performance of the model. GLP guides the generation of SNN outputs by incorporating both global and local characteristics, enhancing the fusion of spatial and temporal features. The TSF is the next influential component. When TSF is removed, our model experiences a performance decrease of 15.1\% in HM and 20.3\% in ZSL. TSF enables our model to dynamically adjust the weights of different time steps based on the output of SNN, improving the efficiency of temporal information extraction. The DTH adjusts the threshold of SNN based on the amount of input information and GLP, which alleviate the spiking noise and improve the robustness of the model. Lastly, the LKC computes the importance of each latent knowledge slot and effectively combines them together.

Furthermore, in Table \ref{TAB2}, we also demonstrate the impact of different loss items on the model performance. We observe that utilizing the complete loss function yields the best HM and ZSL performance across the UCF-GZSL, VGGSound-GZSL, and ActivityNet-GZSL datasets. When both the $\mathcal{L}_{p}$ and $\mathcal{L}_{r}$ losses are removed, our model experiences a decrease of 28.1\% in HM and 57.9\% in ZSL. This experiment verifies the indispensable role of each loss function in the model training process, highlighting the importance of incorporating all loss items to ensure enhanced GZSL and ZSL performance. 

\begin{table}
	\centering
	\begin{threeparttable}
		\renewcommand\arraystretch{1.2}
		\caption{The comparison of different combinations of loss weights.}
		\label{TAB6}
		\setlength{\tabcolsep}{12pt}{
\begin{tabular}{cc|cccc}
\hline \hline
$\alpha$   & $\beta$   & S     & U     & \textit{HM $\uparrow$}    & \textit{ZSL $\uparrow$}   \\ \hline 
0.2 & 0.8 & 55.12 & 20.11 & 29.47          & 25.52          \\
0.8 & 0.2 & 58.13 & 20.89 & 30.74          & 27.16          \\
0.7 & 0.3 & 56.94 & 21.12 & 30.81          & 26.93          \\
0.3 & 0.7 & 54.69 & 21.78 & 31.15          & 28.47          \\ \hline
\rowcolor{lightgray!30}
0.5 & 0.5 & 56.47 & 22.89 & \textbf{32.58} & \textbf{29.72} \\ \hline \hline
\end{tabular}}
\end{threeparttable}
\end{table}

\begin{table}
	\centering
	\begin{threeparttable}
		\renewcommand\arraystretch{1.2}
		\caption{The effectiveness of combining SNN and Transformer.}
		\label{TAB7}
		\setlength{\tabcolsep}{8pt}{
\begin{tabular}{c|cccc}
\hline \hline
Model           & S     & U     & \textit{HM $\uparrow$}    & \textit{ZSL $\uparrow$}   \\ \hline
Transformer+MLP        & 62.43 & 14.31 & 23.28          & 21.03          \\
Spikformer+SNN & 28.12 & 23.69 & 25.72          & 28.54          \\ \hline
\rowcolor{lightgray!30}
Transformer+SNN (STFT)            & 56.47 & 22.89 & \textbf{32.58} & \textbf{29.72} \\ \hline \hline
\end{tabular}}
\end{threeparttable}
\end{table}

\begin{table}
    \centering
    \begin{threeparttable}
        \renewcommand\arraystretch{1.2}
        \caption{The ablation study of the impact of different latent knowledge slots on three benchmark datasets.}
        \label{revised_ablation}
        \setlength{\tabcolsep}{4pt}{
        \begin{tabular}{cccccccc}
        \hline \hline
        \multirow{2}{*}{Number of slots} & \multicolumn{2}{c}{VGGSound-GZSL} & \multicolumn{2}{c}{UCF-GZSL} & \multicolumn{2}{c}{ActivityNet-GZSL} \\ 
        \cmidrule(lr){2-3} \cmidrule(lr){4-5} \cmidrule(lr){6-7}
         & HM & ZSL & HM & ZSL & HM & ZSL \\ 
        \hline
         1 & 9.14 & 6.99 & 28.46 & 29.33 & 13.96 & 11.63 \\
         2 & 9.31 & 7.56 & 30.71 & 29.14 & 13.87 & 11.96 \\
         3 & 9.48 & 8.03 & 32.58 & 30.72 & 14.35 & 12.14 \\
         4 & \textbf{10.06} & \textbf{8.24} & \textbf{31.47} & 30.11 & 15.12 & 12.03 \\
         5 & 9.36 & 8.13 & 31.47 & 30.11 & \textbf{15.38} & \textbf{12.91} \\
         6 & 9.88 & 7.93 & 31.75 &\textbf{ 30.54} & 15.21 & 12.64 \\
        \hline \hline
        \end{tabular}}
    \end{threeparttable}
\end{table}

\subsubsection{The impact of TSF in different time step} The performance variation of the model with and without TSF at different time steps is shown in Fig. \ref{fig3}(a) and \ref{fig3}(d). It is evident that when TSF is introduced, the performance of the model becomes more stable and improved across different time steps. However, there is a slight decrease in both HM and ZSL metrics when the number of time steps increases from 8 to 16. This is because as the number of time steps increases, it leads to increased redundancy in the SNN outputs and significantly higher computational costs. Overall, without TSF has a more significant impact on the ZSL performance, especially in cases of low and high time steps. The TSF enhances the stability and performance of the model across different time steps, providing a more efficient way to combine the outputs of SNN at various time steps and obtain a more comprehensive feature representation.

\subsubsection{The impact of rank constraint in $\boldsymbol{\mathcal{T}}_{c}$}  In Fig. \ref{fig3}(b) and \ref{fig3}(e), we show the impact of different rank constraints in $\boldsymbol{\mathcal{T}}_{c}$ on the model performance at different time steps. A lower rank constraint represents faster inference speed and fewer model parameters, while a higher rank constraint indicates more information and more preserved features. Generally, higher ranks achieve higher accuracy, particularly at lower time steps such as 2 and 4. When the rank is set to 80, the performance of STFT continues to improve as the number of time steps increases, while the other curves show a slight decrease. This improvement may be attributed to the richer information fusion during feature combination. However, when the rank is set to 60, the performance in ZSL is higher than that of rank 80 at time step 8. This is because the output of the SNN is sparse, and a lower rank constraint can filter out redundant features. Considering the overall practical considerations, we believe that selecting a rank of 60 can ensure performance while improving model efficiency.
\subsubsection{The importance of different spiking thresholds} In Fig. \ref{fig3}(c) and \ref{fig3}(f), we demonstrate the impact of dynamic threshold adjustment on the model performance using multimodal and unimodal inputs with different fixed thresholds. Our model is highly sensitive to the spike thresholds of neurons, and as a result, the model performance experiences significant variations with different fixed training thresholds. When using dynamic thresholds, the STFT method outperforms all the fixed threshold methods. Obviously, multimodal inputs yield a great improvement compared to unimodal inputs, highlighting the importance of joint learning. The dynamic threshold adjustment module dynamically adjusts the model threshold by measuring the amount of information in the current input, which leads to enhanced feature representations and highlights the significance of incorporating multimodal inputs for better performance.

\subsubsection{The effectiveness of each loss items} The total loss is formulated as $\mathcal{L}_{all} = \alpha*\mathcal{L}_{t}+\beta*(\mathcal{L}_{p}+\mathcal{L}_{r})$, where $\alpha$ and $\beta$ are the hyperparameters. The additional experiments of the hyperparameters of the loss weights are illustrated in Table \ref{TAB6}. In Table \ref{TAB6}, the equal weights of loss items demonstrated the superiorities compared with other weights combinations.

\subsubsection{The effectiveness of combining SNN and Transformer} In Table \ref{TAB7}, we replace the Transformer to Spikformer (full-spike) and SNN to MLP to certify the effectiveness of the combination of SNN and Transformer. ``Transformer+MLP" (full-float) performs best on the seen classes but has a significant gap with our model on unseen classes and ZSL. The ``Spikformer+SNN" (full-spike) performs struggle in seen classes. The experiment demonstrates the strong domain adaptation abilities of SNN in zero-shot learning. Therefore, combining SNN with Transformer can leverage the characteristics of both types of models and performs the best on the $\mathrm{HM}$ metric.
\begin{table}
	\centering
	\begin{threeparttable}
		\renewcommand\arraystretch{1.2}
		\caption{The statistics for our VGGSound, UCF, and ActivityNet (G)ZSL$^{cls}$ datasets.}
		\label{TAB4}
		\setlength{\tabcolsep}{6pt}{
\begin{tabular}{c|cccc|c}
\hline \hline
\multirow{2}{*}{Dataset} & \multicolumn{4}{c|}{\# Classes} & \#Videos \\ \cline{2-6} 
                         & all   & train     & val(U)   & test(U)   & test(U)     \\ \hline
VGGSound-GZSL$^{cls}$            & 271   & 138   & 69     & 64     & 3200     \\ 
UCF-GZSL$^{cls}$                 & 48    & 30    & 12     & 6      & 845      \\ 
ActivityNet-GZSL$^{cls}$         & 198   & 99    & 51     & 48     & 4052     \\ \hline \hline
\end{tabular}}
\end{threeparttable}
\end{table}

\begin{table}
	\centering
	\begin{threeparttable}
		\renewcommand\arraystretch{1.2}
		\caption{The ablation study on only using Transformer or SNN.}
		\label{revised-tabIX}
		\setlength{\tabcolsep}{10pt}{
\begin{tabular}{c|cccc}
\hline \hline
Model           & S     & U     & \textit{HM $\uparrow$}    & \textit{ZSL $\uparrow$}   \\ \hline
Only Transformer        & 58.41 & 15.23 &   24.16        & 22.31          \\
Only SNN & 26.14 & 22.95 &    24.44       & 27.72          \\ \hline
\rowcolor{lightgray!30}
STFT            & 56.47 & 22.89 & \textbf{32.58} & \textbf{29.72} \\ \hline \hline
\end{tabular}}
\end{threeparttable}
\end{table}

\subsubsection{The impact of different latent knowledge slots}
We show the impact of various latent knowledge slots on model performance in Table \ref{revised_ablation}. These knowledge slots symbolize the latent semantic features present between two modalities. Effectively integrating knowledge slots from diverse modalities helps in discovering and aligning latent cross-modal relationships, facilitating the extraction of stronger multimodal feature representations. We increment the number of latent knowledge slots from 1 to 6 and perform an ablation study on the UCF101, VGGSound, and ActivityNet datasets.

Overall, the performance changes on the VGGSound and ActivityNet datasets are relatively smooth. However, noticeable fluctuations occur on the UCF101 dataset. Table \ref{revised_ablation} shows the performance changes of different latent knowledge slots on VGGSound, where the optimal performance is achieved with four slots. Table \ref{revised_ablation} displays the performance changes on the UCF101 dataset, indicating a general upward trend in the model's performance as the number of slots increases. With three slots, there is an improvement of 14.48\% and 4.7\% on the HM and ZSL datasets, respectively, compared to having just one slot. Table \ref{revised_ablation} also reveals the performance on the ActivityNet dataset, achieving the best performance with five slots.

Having more slots means more latent semantic features, but it can also introduce unnecessary redundancy. Thus, choosing an optimal number of slots is essential for improving model performance.
\subsubsection{The effectiveness of only using Transformer or SNN} Table \ref{revised-tabIX} evaluates the effectiveness of combining SNN and Transformer models. Comparing ``Only Transformer" and ``Only SNN" models, the Transformer-only model excels in performance on seen classes but notably underperforms on unseen classes. Conversely, the SNN-only model struggles with seen data but performs better on unseen classes. This indicates SNN's strong domain adaptation capabilities in zero-shot learning scenarios. The STFT model, which integrates both SNN and Transformer, successfully combines the strengths of both. It achieves the highest score on the HM metric, demonstrating superior overall performance and balance between seen and unseen data. 

\begin{table}
	\centering
	\begin{threeparttable}
		\renewcommand\arraystretch{1.2}
		\caption{The parameter comparison on VGGSound dataset.}
		\label{revised-parameter}
		\setlength{\tabcolsep}{6pt}{
\begin{tabular}{c|cccccc}
\hline \hline
Model           & S     & U     & \textit{HM $\uparrow$}    & \textit{ZSL $\uparrow$}  &\#params &GFLOPS \\ \hline
AVCA \cite{avca}       &14.90	&4.00	&6.31	&6.00	&1.69M	&2.36          \\
AVMST \cite{li2023modality}&14.14	&5.28	&7.68	&6.61	&6.32M	&5.12          \\
MDFT \cite{MDFT} &16.14	&5.97	&8.72	&7.13	&5.51M	&5.62 \\
\hline
\rowcolor{lightgray!30}
STFT           &\textbf{19.22}	&\textbf{6.81}	&\textbf{10.06}	&\textbf{8.24}	&4.16M	&4.27 \\ \hline \hline
\end{tabular}}
\end{threeparttable}
\end{table}

\subsection{Parameter analysis}
As shown in Table \ref{revised-parameter}. ``\#params" represents the number of parameters, ``GFLOPs " represents the computational cost during training. Overall, our model exhibits strong performance in both the number of parameters and computational costs, while ensuring high classification effectiveness. Compared to MDFT, our model reduces the parameters by approximately 32\%, and reduces the GFLOPS from 5.62 to 4.27.

\subsection{Different audio/video extracted networks.}
\begin{table*}
	\centering
	\begin{threeparttable}
		\renewcommand\arraystretch{1.5}
		\caption{We conduct evaluations of STFT along with state-of-the-art (G)ZSL methods on the VGGSound$^{cls}$, UCF$^{cls}$, and ActivityNet$^{cls}$ datasets using features extracted from audio/video classification networks.}
		\label{TAB5}
		\setlength{\tabcolsep}{5pt}{
\begin{tabular}{cccccclcccclcccc}
\hline \hline
\multirow{2}{*}{Type} &
  \multirow{2}{*}{Model} &
  \multicolumn{4}{c}{VGGSound-GZSL$^{cls}$} &
   &
  \multicolumn{4}{c}{UCF-GZSL$^{cls}$} &
   &
  \multicolumn{4}{c}{ActivityNet-GZSL$^{cls}$} \\ \cline{3-6} \cline{8-11} \cline{13-16} 
 &
   &
  S &
  U &
  \textit{HM $\uparrow$} &
  \textit{ZSL $\uparrow$} &
   &
  S &
  U &
  \textit{HM $\uparrow$} &
  \textit{ZSL $\uparrow$} &
   &
  S &
  U &
  \textit{HM $\uparrow$} &
  \textit{ZSL $\uparrow$} \\ \hline
\multirow{2}{*}{ZSL} &
 ALE \cite{ALE} &
  26.13 &
  1.72 &
  3.23 &
  4.97 &
   &
  45.42 &
  29.09 &
  35.47 &
  32.30 &
   &
  0.89 &
  6.16 &
  1.55 &
  6.16 \\
 &
  SJE \cite{SJE} &
  16.94 &
  2.72 &
  4.69 &
  3.22 &
   &
  19.39 &
  32.47 &
  24.28 &
  32.47 &
   &
  37.92 &
  1.22 &
  2.35 &
  4.35 \\
 &
 DEVISE \cite{devise} &
  29.96 &
  1.94 &
  3.64 &
  4.72 &
   &
  29.58 &
  34.80 &
  31.98 &
  35.48 &
   &
  0.17 &
  5.84 &
  0.33 &
  5.84 \\
 &
  APN \cite{APN} &
  6.46 &
  6.13 &
  6.29 &
  6.50 &
   &
  13.54 &
  28.44 &
  18.35 &
  29.69 &
   &
  3.79 &
  3.39 &
  3.58 &
  3.97 \\ \hline
\multirow{5}{*}{\begin{tabular}[c]{@{}c@{}}Audio-visual\\ ZSL\end{tabular}} &
  CJME \cite{cjme} &
  10.86 &
  2.22 &
  3.68 &
  3.72 &
   &
  33.89 &
  24.82 &
  28.65 &
  29.01 &
   &
  10.75 &
  5.55 &
  7.32 &
  6.29 \\
 &
  AVGZSLNet \cite{avgzslnet} &
  15.02 &
  3.19 &
  5.26 &
  4.81 &
   &
  74.79 &
  24.15 &
  36.51 &
  31.51 &
   &
  13.70 &
  5.96 &
  8.30 &
  6.39 \\
 &
  AVCA \cite{avca} &
  12.63 &
  6.19 &
  8.31 &
  6.91 &
   &
  63.15 &
  30.72 &
  41.34 &
  37.72 &
   &
  16.77 &
  7.04 &
  9.92 &
  7.58 \\
 &
  TCaF \cite{tcaf} &
  12.63 &
  6.72 &
  8.77 &
  7.41 &
   &
  67.14 &
  40.83 &
  50.78 &
  44.64 &
   &
  30.12 &
  7.65 &
  12.20 &
  7.96  \\ &
Hyper$^{\mathrm{alignment}}$ \cite{hong2023hyperbolic} &
  12.50 &
  6.44 &
  8.50 &
  7.25 &
   &
  57.13 &
  33.86 &
   42.52 &
  39.80 &
   &
29.77 & 8.77 & 13.55 & 9.13
 \\
 & 
Hyper$^{\mathrm{single}}$ \cite{hong2023hyperbolic}&
12.56 & 5.03 & 7.18 & 5.47 & & 63.47 &34.85 &44.99 &39.86 & & 24.61 &10.10 &14.32 &10.37
 \\
 &
Hyper$^{\mathrm{multiple}}$ \cite{hong2023hyperbolic} &
  15.62 &6.00 &8.67 &7.31 &&74.26 &35.79 &48.30 &52.11 &&36.98 &9.60 &15.25& 10.39 \\
  \hline \rowcolor{lightgray!30}
& STFT (ours) &
  11.74 &
  8.83 &
  \textbf{10.08}&
  \textbf{8.79} &
   &
  61.42 &
  43.81 &
\textbf{51.14}&
  \textbf{49.74} &
   &
  25.12 &
  9.83 &
  \textbf{14.13} &
  \textbf{9.46} \\
\hline \hline
\end{tabular}
}
\end{threeparttable}
\end{table*}
\label{sec:rationale}

\begin{figure}
    \centering
	\includegraphics[scale=0.25]{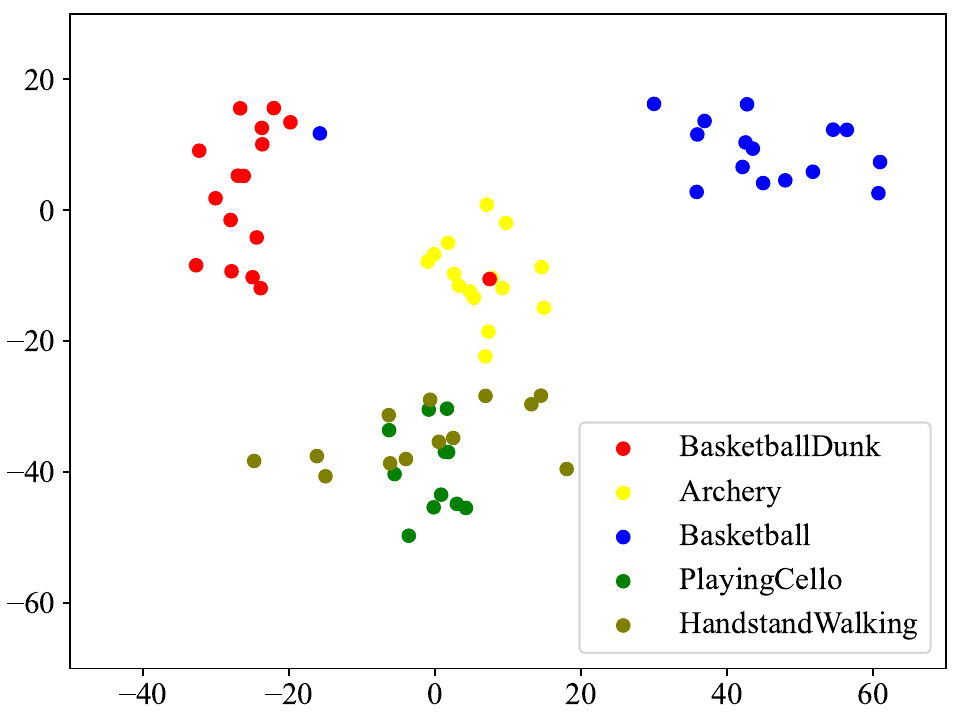}
        \includegraphics[scale=0.25]{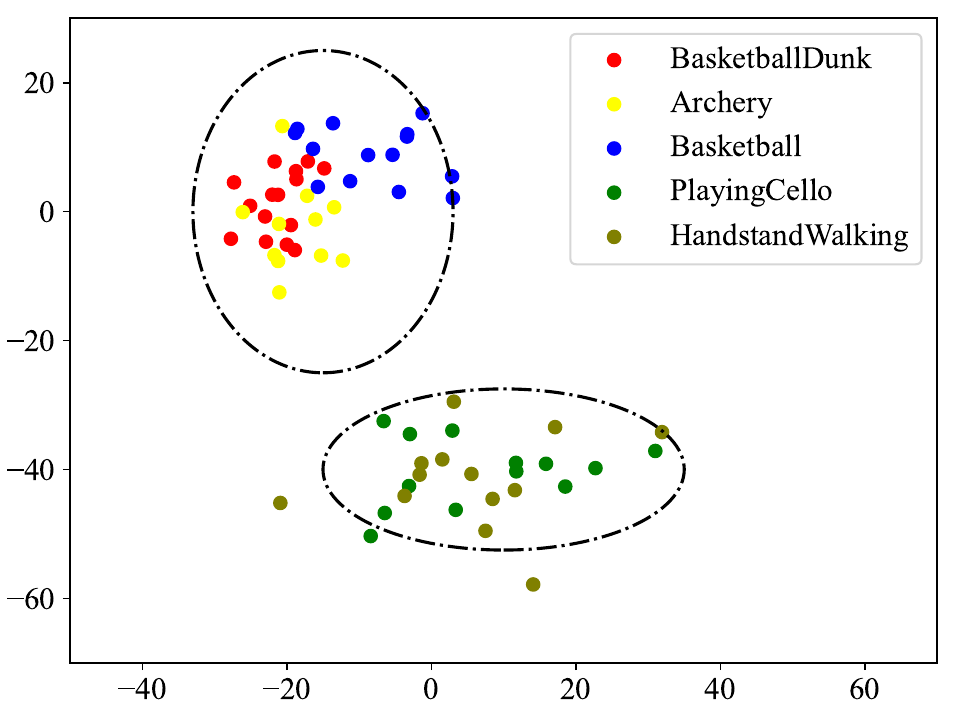}
	\caption{Visualization examples on UCF101. We give t-SNE visualization results for five categories which can be categorized into two parent classes: ``Sports" and ``Instrument".}
 \label{revised_visualization}
\end{figure}

We expand our methodology by incorporating features from audio and video classification networks into our model training and evaluation process. Specifically, we use C3D \cite{C3D}, a network pre-trained on the Sports1M \cite{Sports1M} dataset for video classification, to extract visual features. For audio feature extraction, we use VGGish \cite{VGGish}, pre-trained on the Youtube-8M \cite{youtube8m} dataset. To create a unified feature representation for each video, we average these extracted features over time, resulting in a 4096-dimensional visual feature vector and a 128-dimensional audio feature vector.

To adjust the audio features derived from the Youtube-8M pre-trained network, we make changes to the dataset splits for VGGSound-GZSL, UCF-GZSL, and ActivityNet-GZSL. We remove the test unseen classes that overlap with the Youtube-8M dataset, leading to modified dataset splits named VGGSound-GZSL$^{cls}$, UCF-GZSL$^{cls}$, and ActivityNet-GZSL$^{cls}$. More details on these adjustments are available in Table \ref{TAB4}.

Table \ref{TAB5} presents the comparative results of our STFT model against the baseline models, using the audio and video classification features mentioned above. The STFT model shows exceptional performance across all datasets compared to the baselines. For example, in the VGGSound-GZSL$^{cls}$ dataset, STFT achieves a Harmonic Mean (HM) of 10.08\% and a Zero-Shot Learning (ZSL) accuracy of 8.79\%, outperforming the TCaF model, which records an HM of 8.77\% and a ZSL accuracy of 7.41\%. In the UCF-GZSL$^{cls}$ scenario, STFT reaches an HM of 51.14\% and a ZSL accuracy of 49.74\%, surpassing both AVCA and AVGZSLNet$^{cls}$, which show lower HMs and ZSL accuracies. Likewise, on ActivityNet-GZSL$^{cls}$, AVCA outperforms AVGZSLNet$^{cls}$ in terms of HM and ZSL accuracy. These results highlight STFT's consistent superiority over competing models, attributing this success to the innovative integration of our temporal-semantic tucker fusion module, which effectively combines SNN and Transformer outputs for improved multi-scale fusion and performance.

\begin{figure}
    \centering
	\includegraphics[scale=0.4]{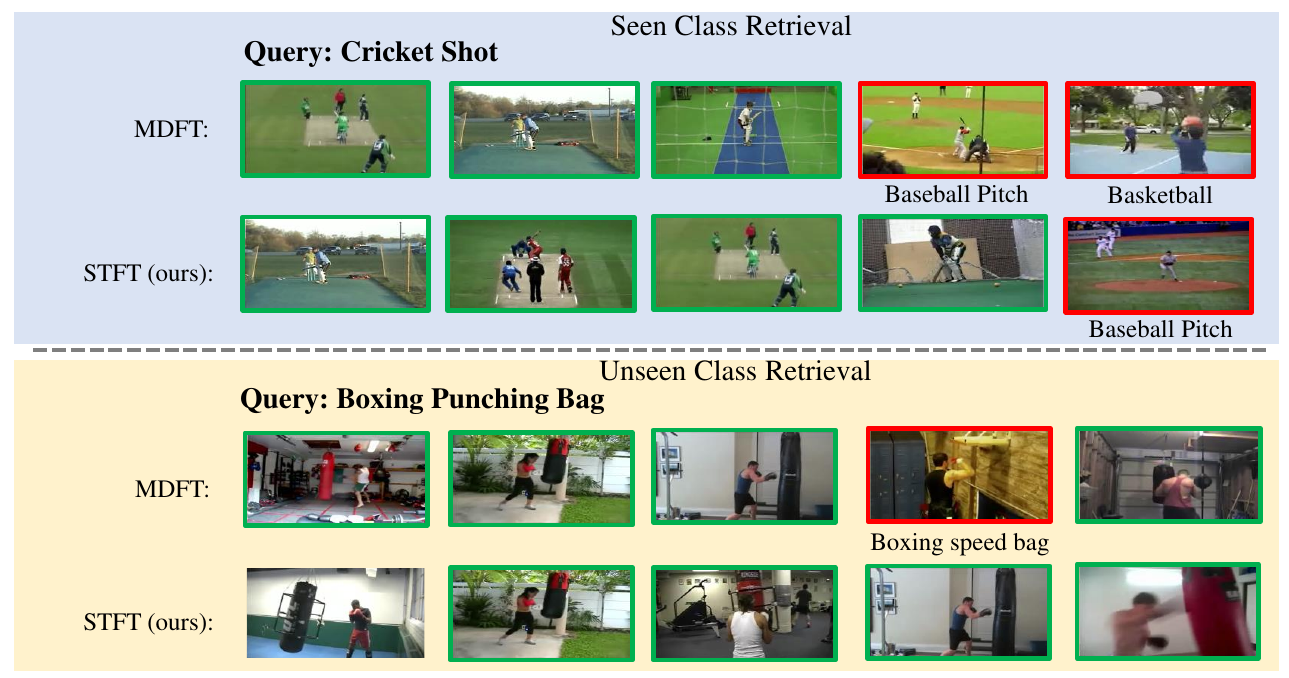}
	\caption{Qualitative comparison results compared with MDFT.}
 \label{fig5}
\end{figure}
\subsection{Visualization results}
We use t-SNE visualization to demonstrate the advantages of the proposed STFT in exploring intrinsic correlations within multimodal data, as shown in the Fig. \ref{revised_visualization}. In the UCF101 dataset, STFT actively clusters features from the same parent category and separates features from different parent categories. For example, features such as ``basketball" and ``basketballdunk," which belong to the same parent category ``sport," are brought closer, while features like ``PlayingCello" and ``HandstandWalking" from the ``instrument" category are separated. These visualizations illustrate how our method explores correlations between different types of data.

\subsection{Qualitative Comparison}
We demonstrate the qualitative comparison results with recent SOTA method MDFT in Fig. \ref{fig5}. MDFT focus on decoupling motion and background information, while our model coupling the outputs of SNN and Transformer effectively. In this paper, we address the challenges of time steps and spiking redundancy in SNN, and the output heterogeneity between SNN and Transformer. In sports classes with frequent changes in motion information, STFT demonstrate superiorities compared with MDFT due to the less spiking redundancy.

\subsection{Limitations} Although our model has demonstrated SOTA performance in HM on three benchmark datasets, we observed a slight decrease on ZSL in UCF101. This may caused by the fixed rank constraint. A potential solution could be dynamically setting the rank constraint of the current temporal-semantic tucker fusion based on the singular values of the input information. This adaptive adjustment strategy of rank constraint could potentially improve the ZSL performance.

\subsection{Scalability Discussion}
The proposed Spiking Tucker Fusion Transformer (STFT) is designed for scalability, effectively handling larger datasets and complex audio-visual sequences. The STFT uses a temporal-semantic fusion module based on Tucker decomposition, enabling multi-scale fusion of SNN and Transformer outputs. This design ensures the number of parameters remains manageable, maintaining computational efficiency even with larger datasets. Efficient data loading and batching strategies are used to handle larger datasets, ensuring memory constraints are not exceeded and performance is maintained. Additionally, the STFT adapts to different audio-visual sequence complexities through dynamic adjustments, such as the TSF for synthesizing temporal information and GLP for reducing spike noise and enhancing robustness. These components enable the model to effectively manage sequence complexity, ensuring robust performance across diverse scenarios. Overall, the STFT's design and components enhance its scalability and applicability in real-world scenarios involving large datasets and complex audio-visual sequences.

\section{Conclusion}
In conclusion, this paper introduces the Spiking Tucker Fusion Transformer (STFT) model for audio-visual zero-shot learning. The STFT model effectively combines Spiking Neural Networks (SNNs) and Transformers, integrating both temporal and semantic information to generate robust representations. By introducing time-step factors (TSF), the significance of each time step in influencing the SNN's output is dynamically measured, leading to improved performance. To guide the formation of input membrane potentials and reduce spike noise, a global-local pooling (GLP) method is proposed. Additionally, the thresholds of the spiking neurons are adjusted dynamically based on semantic and temporal cues, enhancing the model's robustness. We propose a temporal-semantic tucker fusion module to achieves multi-scale fusion of SNN and Transformer outputs while maintaining full second-order interactions. The experimental results demonstrate that the proposed STFT model outperforms existing methods in audio-visual zero-shot learning tasks.



 



\begin{IEEEbiography}[{\includegraphics[width=1in,height=1.25in,clip,keepaspectratio]{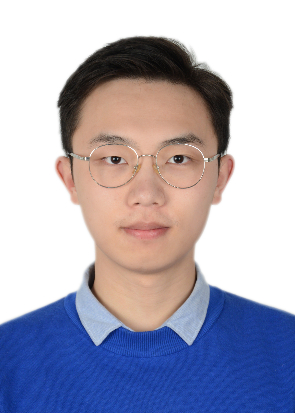}}]{Wenrui Li} received the B.S. degree from the School of Information and Software Engineering, University of Electronic Science and Technology of China (UESTC), Chengdu, China, in 2021. He is currently working toward the Ph.D. degree from the School of Computer Science, Harbin Institute of Technology (HIT), Harbin, China. His research interests include multimedia search, joint source-channel coding, and spiking neural network. He has authored or co-authored more than 15 technical articles in referred international journals and conferences. He also serves as a reviewer for IEEE TCSVT, IEEE TMM, NeurIPS, ECCV, AAAI, and ACM MM.
\end{IEEEbiography}

\begin{IEEEbiography}[{\includegraphics[width=1in,height=1.25in,clip,keepaspectratio]{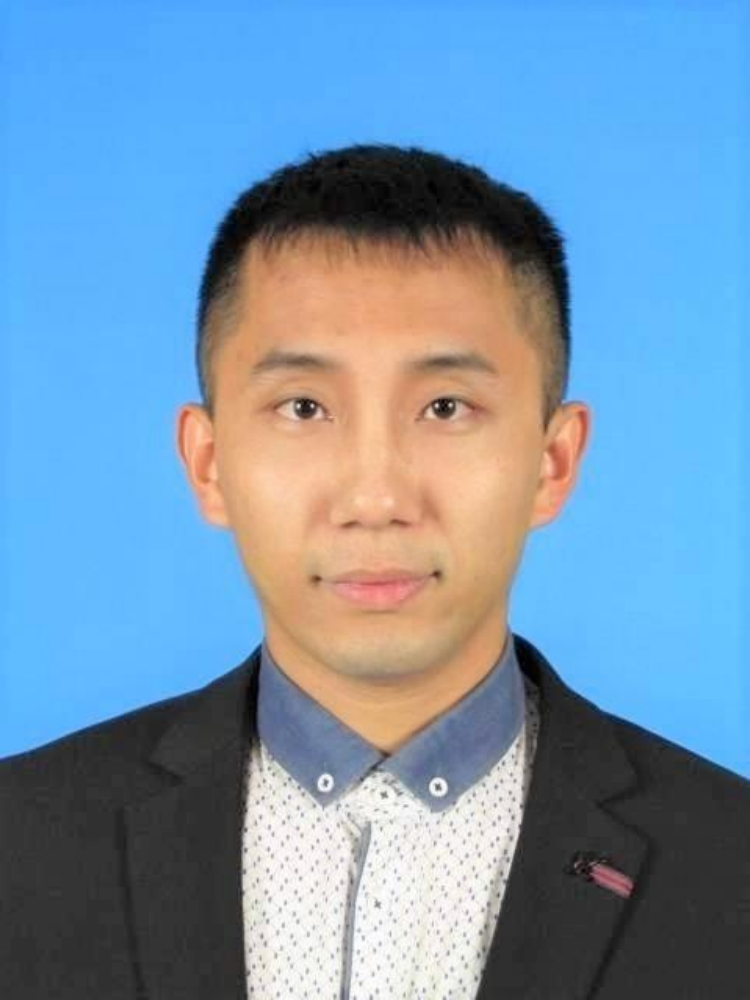}}]{Penghong Wang} received the M.S. degree in computer science and technology from Taiyuan University of Science and Technology, Taiyuan, China, in 2020. He is currently pursuing the Ph.D. degree with the School of Computer Science, Harbin Institute of Technology, Harbin, China. His main research interests include wireless sensor networks, joint source-channel coding, and computer vision. He has authored or co-authored more than 20 technical articles in referred international journals and conferences. He also serves as a reviewer for IEEE TVT, IEEE TAES, IEEE CE, IEEE IOTJ, NeurIPS, ECCV, AAAI, and ACM MM.
\end{IEEEbiography}

\begin{IEEEbiography}[{\includegraphics[width=1in,height=1.25in,clip,keepaspectratio]{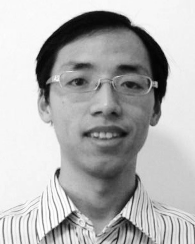}}]{Ruiqin Xiong}
(Senior Member, IEEE) received the B.S. degree in computer science from the University of Science and Technology of China, Hefei, China, in 2001, and the Ph.D. degree in computer science from the Institute of Computing Technology, Chinese Academy of Sciences, Beijing, China, in 2007. 
From 2002 to 2007, he was a Research Intern with Microsoft Research Asia.
From 2007 to 2009, he was a Senior Research Associate with the University of New South Wales, Sydney, NSW, Australia. 
In 2010, he joined the School of Electronic Engineering and Computer Science, Peking University, Beijing, where he is currently a Professor. 
He has authored or coauthored more than 140 technical articles in referred international journals and conferences.
His research interests include image and video processing, statistical image modeling, deep learning, neuromorphic camera, and computational imaging.
He was a recipient of the Best Student Paper Award from the SPIE Conference on Visual Communications and Image Processing in 2005 and the Best Paper Award from the IEEE Visual Communications and Image Processing in 2011.
He was a co-recipient of the Best Student Paper Award from the IEEE Visual Communications and Image Processing in 2017.
\end{IEEEbiography}

\begin{IEEEbiography}[{\includegraphics[width=1in,height=1.25in,clip,keepaspectratio]{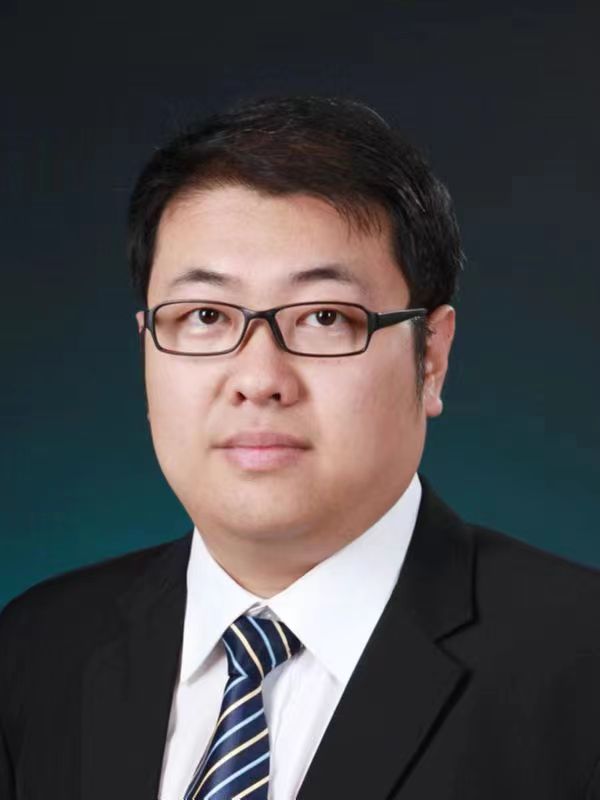}}]{Xiaopeng Fan} (Senior Member, IEEE) received the B.S. and M.S. degrees from the Harbin Institute of Technology (HIT), Harbin, China, in 2001 and 2003, respectively, and the Ph.D. degree from the Hong Kong University of Science and Technology, Hong Kong, in 2009. In 2009, he joined HIT, where he is currently a Professor. From 2003 to 2005, he was with Intel Corporation, China, as a Software Engineer. From 2011 to 2012, he was with Microsoft Research Asia, as a Visiting Researcher. From 2015 to 2016, he was with the Hong Kong University of Science and Technology, as a Research Assistant Professor. He has authored one book and more than 170 articles in refereed journals and conference proceedings. His research interests include video coding and transmission, image processing, and computer vision. He was the Program Chair of PCM2017, Chair of IEEE SGC2015, and Co-Chair of MCSN2015. He was an Associate Editor for IEEE 1857 Standard in 2012. He was the recipient of Outstanding Contributions to the Development of IEEE Standard 1857 by IEEE in 2013.
\end{IEEEbiography}

\end{document}